\documentclass[12pt]{article}
\usepackage[margin = 1in]{geometry}
\usepackage{amsmath, amsfonts, amssymb, amsthm}

\RequirePackage[authoryear]{natbib}

\usepackage{hyperref}
\usepackage[plain,noend]{algorithm2e}
\usepackage{url}

\usepackage{titlesec}

\newcommand{\indep}{\perp \!\!\! \perp}

\newcommand{\E}{\mathbb E}

\newtheorem{assumption}{Assumption}
\newtheorem{proposition}{Proposition}
\newtheorem{corollary}{Corollary}[proposition]
\RequirePackage{bm,makecell,pdflscape,multirow}
\usepackage{tikz}
\usepackage{varwidth}
\usetikzlibrary{arrows.meta}
\usepackage{cancel}
\usepackage{makecell}
\usetikzlibrary {shapes.multipart}
\usepackage{yhmath}
\usetikzlibrary{shapes,arrows,fit,calc,positioning,automata}
\usepackage{afterpage}
\usepackage{authblk}
\usepackage{setspace}
\titleformat*{\section}{\large\bfseries}
\titleformat*{\subsection}{\normalsize\bfseries}

\begin{document}
\title{An Instrumental Variables Framework to Unite Spatial Confounding Methods}

\author[1]{Sophie M. Woodward}
\author[1,2]{Mauricio Tec}
\author[1]{Francesca Dominici}
\affil[1]{Biostatistics, Harvard University}
\affil[2]{Computer Science, Harvard University}
\date{November 2025}

\maketitle
\begin{abstract}
Studies investigating the causal effects of spatially varying exposures on outcomes often rely on observational and spatially indexed data. A prevalent challenge is unmeasured spatial confounding, where an unobserved spatially varying variable affects both exposure and outcome, leading to biased estimates and invalid confidence intervals. There is a very large literature on spatial statistics that attempts to address unmeasured spatial confounding bias; most of this literature is not framed in the context of causal inference and relies on strict assumptions. 
In this paper, we propose an instrumental variables (IV) framework that unifies and extends existing methods for addressing unmeasured spatial confounding bias. This framework reveals that many spatial confounding methods can be viewed as IV methods, in which small-scale spatial variation in exposure operates as the instrument, providing a common theoretical foundation for approaches that previously appeared distinct. The framework clarifies that these methods share a common set of assumptions and differ primarily in how small-scale spatial variation is defined, while offering a general strategy for constructing instruments. It also extends to identify and estimate a broad class of causal effects, including the exposure response curve, without requiring a linear outcome model. We apply our methodology in simulation and to a national data set of 33,255 zip codes to estimate the effect of enforcing air pollution exposure levels below $6$--$12\mu g/m^3$ on all-cause mortality while adjusting for unmeasured spatial confounding. 
\end{abstract}

\newpage
\doublespacing
\section{Introduction}

Unmeasured spatial confounding is a central challenge in causal inference studies that rely on observational spatial data. We denote by $U$ an unmeasured confounder that exhibits spatial autocorrelation; that is, two observations in close spatial proximity are highly correlated and their correlation decays as the distance between them grows. Failing to adjust for $U$ results in biased estimates of causal effects and invalid confidence intervals \citep{fewell2007impact, robins2000sensitivity, tec2024spatial,   vanderweele2011bias}. However, an important aspect of an unmeasured confounder with spatial autocorrelation is that spatial information can be leveraged to adjust for confounding bias, provided that the confounder varies smoothly with space
\citep{gilbert2021causal}.

The spatial confounding literature emphasizes the importance of spatial scale of both the unmeasured confounder and the exposure of interest. \cite{paciorek2010a} examined the bias of the estimated exposure coefficient from both spatial random effects and penalized spline models, assuming a linear outcome model in which Gaussian processes with Mat\'ern correlation generate both confounder and exposure. The author found that bias can be reduced only if the unmeasured spatial confounder varies more smoothly across space than the exposure, with smoothness defined by the spatial range parameter of the Mat\'ern correlation. Although \cite{khan2023re, narcisi2024effect} showed that these findings are sensitive to distributional assumptions, this result has inspired a line of work on spatial confounding that recommends exploiting small-scale or non-spatial variation in exposure to estimate the statistical parameter of interest \citep{bobb2022accounting, dupont2022a, dupont2023demystifying,  giffin2021instrumental, gilbert2023consistency,guan2022spectral, hanks2015restricted, keller2020a, marques2022mitigating, nobre2021effects,    prim2025spectral,thaden2018structural,
  urdangarin2024simplified,  wiecha2024two}.

Despite the extensive body of research on spatial confounding, the field remains conceptually fragmented. The literature presents multiple definitions of spatial confounding and spatial scale \citep{donegan2024plausible, gilbert2021causal, khan2023re}, and existing methods rely upon ad hoc formulations of the same underlying intuition, that small-scale spatial variation in exposure can be leveraged to estimate the exposure effect, without a shared theoretical foundation. Moreover, prior work has largely focused on identifying the exposure coefficient in a linear model, rather than nonparametric identification of more general causal effects such as the exposure response curve. An exception is \cite{gilbert2021causal}, who establish nonparametric identification under the assumptions that (i) the unmeasured confounder is a continuous function of spatial coordinates and (ii) the exposure exhibits nonspatial variation. In summary, 
the lack of coherence in the literature obscures connections among approaches, complicates the comparison of their assumptions, and impedes progress toward a unified theory of causal identification.

To address this fragmentation, this paper introduces a new framework that reformulates several existing spatial confounding methods as IV methods. Specifically, our framework offers the following new insights. 

\begin{enumerate}
\item Many spatial confounding methods can be viewed as IV methods, where small-scale spatial variation corresponds to an IV. By situating several existing approaches within a unified theoretical framework, we demonstrate that they depend on a common set of five core assumptions. The primary distinction between methods lies in the form of small-scale exposure variation that is presumed to be uncorrelated with the unmeasured confounder. This unification links the spatial confounding literature with the literature on IV methods, potentially opening several new avenues of research.

\item The framework provides a general strategy for constructing an IV: choose any spatial basis that represents the confounded component of exposure, project the exposure onto this basis, and use the orthogonal residual as the IV. In this view, existing spatial confounding methods differ only by the chosen basis, and the same procedure yields new estimators via alternative spatial bases.

\item The framework provides a mathematical formulation of the identification requirement: the exposure must vary outside the spatial basis associated with the unmeasured confounder. 

\item The framework can be extended further to allow for the identification and estimation of a more general class of causal effects, including the exposure response curve, without requiring a linear outcome model. To our knowledge, this is the first methodology to combine flexible causal effect estimation with assumptions about the spatial basis of the unmeasured confounding.
\item These results apply more generally beyond the settings of unmeasured spatial confounding. Our framework shows that identification requires the exposure to additively decompose into two components: one that is conditionally dependent on the unmeasured confounder and one that is not. In principle, these components need not be defined by a spatial basis. The framework clarifies that any source of unconfounded variation can fulfill this role, making the identification strategy broadly applicable.

\end{enumerate}

To illustrate the practical implications of our proposed methodology, we conduct simulations and a data application focusing on a truncated exposure effect that quantifies the impact of restricting exposure levels to lie below a predetermined cutoff. This causal effect is appealing as it depends on a relatively weak positivity assumption and bears clear policy relevance. In the data application, we analyze a national data set of 33,255 ZIP codes to estimate the effect of enforcing air pollution exposure levels below $6$--$12\mu g/m^3$ during 2001--2010 on all-cause mortality in the Medicare population during 2011--2016, leveraging small-scale spatial variation in air pollution exposure as an instrument. In both the simulation and data application, our approach successfully reduces bias from omitted confounders, and its performance improves further when combined with \citet{gilbert2021causal}, suggesting a promising synergy between the two approaches.

\section{An IV framework that unites spatial confounding methods}
\label{framework}
\subsection{Spatial+ as an instrumental variables method}

In this section, we show that several spatial confounding methods can be viewed as IV methods where small-scale spatial variation in exposure operates as an IV. An IV's validity rests on the following assumptions: 1) relevance, meaning that the IV influences the exposure; 2) exogeneity, meaning that the IV is independent of unmeasured confounders conditional on covariates; and 3) exclusion restriction, meaning that the IV only affects the outcome through the exposure \citep{angrist1995identification, baiocchi2014instrumental}.

Before presenting our framework, we illustrate its key components and build intuition using spatial+, a spatial confounding method that has received considerable attention in recent years \citep{dupont2022a}. Let $A = (A_1, \ldots, A_n)^T$, $Y = (Y_1, \ldots, Y_n)^T$, 
$U = (U_1, \ldots, U_n)^T,$ denote the exposure, response, and unmeasured confounder, respectively, measured for $n$ units with corresponding spatial coordinates $S_1, \ldots, S_n$. This notation accommodates both geostatistical data (randomly sampled $S_i$) and areal data (fixed $S_i$, e.g., areal centroids). Measured confounders are omitted for simplicity.

The data generation process for spatial+ assumes that 
\begin{align*}
    A_i &= g(S_i) + \epsilon_i^A,\\
    Y_i &= \beta_0 + \beta A_i + f(S_i) + \epsilon_i^Y
\end{align*} where $f,g$ are unknown, fixed, bounded functions. This scenario exemplifies unmeasured spatial confounding, because the unobserved function $f(S)$ represents an unmeasured spatial confounder: it directly influences the outcome, correlates with the exposure through $g$, and exhibits spatial autocorrelation. An ordinary least squares regression of $Y$ on $A$, without accounting for $S$, will result in a biased estimate of $\beta$ due to the correlation between the exposure $A_i$ and $\epsilon_i = f(S_i) + \epsilon_i^Y$. Spatial+ mitigates this spatial confounding bias via a two-stage procedure. 

First, regress the exposure $A$ on a smooth function of spatial coordinates $S$ (typically a thin plate spline) to obtain the fitted values $\hat{g}$ and residuals $A - \hat{g}(S)$. Second, regress outcome $Y$ on first-stage residuals $A - \hat{g}(S)$ and a thin plate spline of spatial coordinates $S$. The dimension of the thin plate spline basis used in this stage is identical to that of stage 1. The estimated coefficient of $A - \hat{g}(S)$ yields the final estimate of the statistical parameter $\beta$.

Here is the crucial observation: this method's validity relies on the assumption that the estimated residual $A - \hat{g}(S)$ of the first stage regression is not correlated with the unobserved spatial function $f(S)$, which represents the unmeasured confounder. To see this, observe that the resulting estimate of $\beta$ from stage 2 converges in probability to \begin{align*}
    \frac{\text{cov}(Y, A - \hat{g}(S))}{\text{var}(A - \hat{g}(S))} &=  \frac{\text{cov}(\beta_0 + \beta A + \epsilon, A - \hat{g}(S))}{\text{var}(A - \hat{g}(S))} = \beta + \frac{ \text{cov}(\epsilon, A - \hat{g}(S))}{\text{var}(A - \hat{g}(S))} = \beta,
\end{align*}where the last equality holds if and only if $A - \hat{g}(S)$ is uncorrelated with $\epsilon$, or equivalently, if $A - \hat{g}(S)$ is uncorrelated with $f(S)$. It is also required that $\text{var}(A - \hat{g}(S)) \neq 0$, i.e., the exposure $A$ cannot be collinear with the spatial basis.

This observation demonstrates that spatial+ can be viewed as an instrumental variables method, in which $A - \hat{g}(S)$ corresponds to the instrument. First, $A - \hat{g}(S)$ is relevant, since $\text{cov}(A, A - \hat{g}(S)) = \text{var}(A - \hat{g}(S))$. Second, it is assumed that $A - \hat{g}(S)$ is exogeneous, since the validity of unsmoothed spatial+ relies on $\text{Cor}(A - \hat{g}(S), \epsilon) = \text{Cor}(A - \hat{g}(S), f(S)) = 0$. Third, $A - \hat{g}(S)$ obeys exclusion restriction, which is implicit in the functional form of the data-generating process. In fact, unsmoothed spatial+ is equivalent to two-stage least squares, a technique commonly used in IV settings, since an identical estimate of $\beta$ could be obtained simply by regressing $Y$ on the projection of $A$ onto the instrument $A - \hat{g}(S)$ in the second stage. Without smoothing, this projection coincides with the instrument itself, $A - \hat{g}(S)$.

There are several takeaways here that motivate a general framework that unites spatial confounding methods under a common set of assumptions. Unsmoothed spatial+ relies on the assumption that the exposure can be additively decomposed into two random variables, $\hat{g}(S)$ and $A - \hat{g}(S)$, such that $\hat{g}(S)$ is correlated with the unmeasured confounder and $A - \hat{g}(S)$ is uncorrelated with the unmeasured confounder. Furthermore, $A - \hat{g}(S)$, which can be viewed as an instrument, must have non-zero variance to ensure the identifiability of $\beta$. In the context of spatial+, this means that the exposure must exhibit variation that is not spanned by the spatial basis that is used in the thin-plate spline regressions. 

\subsection{An IV framework for linear outcome models}
We now formally introduce our IV framework for spatial confounding methods assuming linear outcome models; this framework is extended to more general outcome models in the subsequent section. This framework encompasses six existing spatial confounding methods as special cases, including spatial+, and further clarifies the distinct underlying assumptions and estimation strategy that each method employs \citep{dupont2022a, guan2022spectral, keller2020a, 
thaden2018structural,
urdangarin2024simplified, wiecha2024two}. A detailed mapping of each method to our framework is provided in the Supplementary Material.

All methods share five fundamental assumptions based on an implicit decomposition of exposure into confounded and unconfounded components. As shown in Table \ref{tab:connections}, the explicit values of these components, $A_{C}$ and $ A_{UC}$, is what differentiates each method. The assumptions are:

\begin{assumption}[Linear outcome model]\label{a1}
    $Y_i = \beta_0 + \beta A_i + \epsilon_i$
\end{assumption}
\begin{assumption}[Additive decomposition of exposure]\label{b1}
    $A_i = A_{C_i} + A_{UC_i}$ 
\end{assumption}
\begin{assumption}[$A_{UC}$ uncorrelated with $A_{C}$]\label{c1}
    $A_{C_i} \perp A_{UC_i}$
\end{assumption}
\begin{assumption}[$A_{UC}$ uncorrelated with error] \label{d1}
    $\epsilon_i \perp A_{UC_i}$
\end{assumption} 
\begin{assumption}[$A_{UC}$ has nonzero variance]\label{e1}
    $\text{var}(A_{UC}) > 0$
\end{assumption}
\noindent where $\perp$ denotes orthogonality (zero correlation). Here, $A_{C}$ and $ A_{UC}$ are two random variables whose sum equals the exposure $A$. $A_{C}$ and $A_{UC}$ are correlated and uncorrelated, respectively, with the spatial error $\epsilon$. In spatial+, we had $A_{C} = \hat{g}(S)$ and $A_{UC} = A - \hat{g}(S)$. To build intuition motivating Section \ref{casestudies}, imagine 
$A$ as long-term average air pollution, $A_C$ as regional background pollution from economic activity and traffic, and $A_{UC}$ as finer-scale pollution variation arising from wildfire plumes, local winds, or terrain. Finally, although typically omitted from the data-generating model, the unmeasured confounder may be encoded in the error $\epsilon$, for example $\epsilon_i = U_i + \epsilon_i^Y$ where $\epsilon_i^Y$ is i.i.d. exogeneous error. 

\begin{table}
\renewcommand{\arraystretch}{1.35}
{\begin{tabular}{p{2.5cm}p{5.75cm}lc}
    {Paper} &  \makecell[l]{Spatial basis or method \\used to obtain $A_C$}  
    & \makecell[l]{The instrument \\$A_{UC} = A-A_C$}& {Method} 
    \\
    \hline
        \cite{dupont2022a} & \makecell[l]{thin-plate spline  basis} 
        & $A-\hat{g}(S)$
        & 2SLS 
        \\
        \cite{urdangarin2024simplified} 
        & \makecell[l]{$k+1$ eigenvectors $v_{n-k}, \ldots, v_{n}$ \\ of the spatial precision  matrix}
        & $A-\sum_{i = n-k}^{n}v_i  v_i^T A$   & 2SLS 
        \\      
        \cite{keller2020a} & \makecell[l]{Fourier/wavelet/thin plate spline \\ basis of dimension $m$, $H_m$}
        & \makecell[l]{$A-H_m(H_m^T H_m)^{-1} H_m^T A$  
        }  
        & 2SRI
        \\      
        \cite{guan2022spectral} & \makecell[l]{$n-1$ eigenvectors $v_1, \ldots, v_{n-1}$ \\ of the Graph  Laplacian }
        & \makecell[l]{$A - \sum_{k = 1}^{n-1} v_k v_k^T A$ }  & 2SRI 
        \\        
        \cite{thaden2018structural} & \makecell[l]{$d$ region indicators $z_{1}, \ldots, z_{d}$ } 
        & \makecell[l]{$A - \sum_{k = 1}^d z_{k}\gamma_{1k}$ }
         & double pred. 
        \\
        \cite{wiecha2024two} & universal kriging  
        & \makecell[l]{$A - \hat{g}(S)$ }  & double pred. 
    \end{tabular}}
    \label{tab:connections}
    \caption{2SLS, two-stage least squares; 2SRI, two-stage residual inclusion; double pred., double prediction. ``Spatial basis or method used to obtain $A_C$" refers to the basis used to decompose the exposure into $A = A_{C} + A_{UC}$. See the Supplementary Material for further details and notation.}
\end{table}

We argue that $A_{UC}$ is an instrumental variable in the following way. First, $A_{UC}$ is relevant: this is encoded in Assumptions \ref{b1}, \ref{c1}, and \ref{e1}, which imply that $\text{cov}(A, A_{UC}) = \text{cov}(A_{C} + A_{UC}, A_{UC}) = \text{var}(A_{UC}) > 0$. Second, $A_{UC}$ is exogeneous, in the sense that $\epsilon \perp A_{UC}$ by Assumption \ref{d1}. Third, $A_{UC}$ obeys exclusion restriction. This property is implicit in the form of the outcome model imposed by Assumption \ref{a1}.   

Our framework distinguishes each method along two lines. First, each method specifies a different decomposition of exposure into components correlated and uncorrelated with the spatial error, determined by the choice of spatial basis. The correlated component $A_C$ is obtained by projecting the exposure onto this basis, and the uncorrelated component $A_{UC}$ is the corresponding residual; this construction directly enforces Assumptions \ref{b1} and \ref{c1}. Methods differ in the type of basis (e.g., Graph Laplacian eigenvectors or splines) and in its dimension. These choices, in turn, determine the requirements for identification of the exposure coefficient $\beta$: identification requires variation in exposure that is not spanned by the spatial basis defining $A_{C}$, so that the instrument $A_{UC}$ has nonzero variance.

The second criterion we use to classify spatial confounding methods within our framework is the IV approach used to leverage the instrument $A_{UC}$. Each of the six methods employs one of the following approaches for the estimation of $\beta$: two-stage least squares (2SLS), where the outcome is regressed on the projection of exposure onto the instrument, which coincides with $A_{UC}$ itself; two-stage residual inclusion (2SRI), where the outcome is regressed on both $A_{UC}$ and $A_{C}$; or double prediction, where the spatial variation used to construct $A_{C}$ is first regressed away from the outcome, followed by regressing the resulting residuals on $A_{UC}$. In fact, we establish the following result: under Assumptions \ref{a1}--\ref{e1}, the coefficient estimates obtained by 2SLS, 2SRI, or double prediction are equivalent, and they converge in probability to $\text{cov}(Y, A_{UC})/\text{var}(A_{UC}) = \beta$. Consequently, if small-scale spatial variation in exposure is successfully extracted and uncorrelated with the spatial error, then 2SLS, 2SRI, and double prediction can each consistently recover the true parameter of interest, $\beta$.

We conclude with three new insights that emerge from our framework. First, our theoretical unification provides a general strategy for constructing an IV: choose any spatial basis that represents the confounded component of exposure, project the exposure onto this basis, and use the orthogonal residual as the IV. From this viewpoint, existing spatial confounding methods differ only by the chosen basis, and the alternative bases naturally yield new spatial confounding estimators. Importantly, however, the IV's validity is basis-dependent; the residualized small-scale spatial variation may be uncorrelated with the unmeasured confounder (satisfying Assumption 4) only under a specific basis, which in turn affects the method's validity---a point also raised by \cite{keller2020a}. This explains why some spatial confounding methods are unbiased in settings where others are not.

Second, the spatial variation assumed to be correlated with the unmeasured confounder directly determines how the exposure must vary to ensure identification. We formalize this in Assumption \ref{e1}, $\text{var}(A_{UC}) > 0$; this condition is closely related to the positivity assumption in Proposition 1 of \cite{gilbert2023consistency}. For example, if state-level variation in air pollution is assumed to be correlated with unmeasured confounders, identification requires the presence of additional uncorrelated variation in air pollution, such as within-state variation. This reformulation also implies that identification can be achieved even when the exposure (i) is perfectly smooth in space, by specifying which spatial scales are confounded versus not \citep{gilbert2021causal}, or (ii) appears smoother than the unmeasured confounder, provided some component of variation remains independent, e.g. large-scale variation \citep{paciorek2010a}. We explore both of these scenarios in the simulation study in Section \ref{simulation}. 

Third, this identification strategy has potential applications beyond spatial confounding contexts. Assumptions \ref{a1}--\ref{e1} enable consistent estimation of $\beta$, regardless of how $A_{UC}$ and $A_C$ are defined. Identification simply requires the exposure to additively decompose into two components: one that is correlated with the unmeasured confounder and one uncorrelated with it. The uncorrelated component need not be tied to small spatial scales. While small-scale spatial variation is often invoked for scientific reasons, any source of unconfounded variation can fulfill this role, making the framework broadly applicable in a wide range of contexts. In the Discussion, we describe other settings where this identification strategy may be feasible. 

\section{Extending the IV framework to estimate more general causal effects}
\label{erf}
The spatial confounding methods in Section \ref{framework} rely on a linearity assumption (Assumption \ref{a1}), which severely limits their applicability in practice. Many scientific questions require estimating more flexible causal effects, such as a potentially non-linear exposure response curve or the effects of interventions that modify the value of exposure based on the observed exposure and covariates. In this section, we advance the IV framework introduced in Section \ref{framework}, demonstrating that it can be extended to identify and estimate the effects of these types of causal effects in the presence of unmeasured spatial confounding. The central theoretical insight is that controlling for measured covariates and a smooth spatial trend in exposure renders the exposure itself conditionally independent of potential outcomes, thereby enabling the identification of many causal effects.

Let $A = (A_1, \ldots, A_n)^T \in \mathbb R^n$, $Y = (Y_1, \ldots, Y_n)^T \in \mathbb R^n$, $X = (X_1, \ldots, X_n)^T \in \mathbb R^{n \times p}$,
$U = (U_1, \ldots, U_n)^T \in \mathbb R^{n \times q},$ denote the exposure, response, measured confounders, and unmeasured confounders, respectively, measured for $n$ units with the corresponding spatial coordinates $S_1, \ldots, S_n$. We introduce here a scalar exposure and outcome but allow for an arbitrary number of measured and unmeasured confounders. Further denote $Y_i(a)$ as the potential outcome for unit $i$ under exposure value $a$. We replace Assumptions \ref{a1}--\ref{e1} of the previous section with the following four assumptions:

\begin{assumption}[Consistency]\label{a}
    $Y_i = Y_i(a)$ when $A_i = a$.
\end{assumption}
\begin{assumption}[Unspecified outcome model] \label{b}
    $Y_i(a) = m(a, X_i, U_i)$.
\end{assumption}
\begin{assumption}[Additive decomposition of exposure]\label{c}
    $A_i = A_{UC_i}+ A_{C_i}$ for some  
    random variables $A_{C} = (A_{C_1}, \ldots, A_{C_n})^T$, $A_{UC} = (A_{UC_1}, \ldots, A_{UC_n})^T$, with $\text{var}(A_{UC})>0$. 
\end{assumption}

\begin{assumption}[Conditional independence of instrument]\label{d}
    $A_{UC_i} \indep (U_i, A_{C_i}) \mid X_i$.
\end{assumption}

Assumptions \ref{a}--\ref{d} are represented in the causal graph in Figure \ref{fig:dag}. These assumptions build upon the IV theory of \cite{imbens2009identification}, although their results were not developed in the context of unmeasured spatial confounding. See the Supplementary Material for additional details.

Assumption \ref{a} is consistency. In particular, it entails the no-interference component of the Stable Unit Treatment Value Assumption (SUTVA): unit $i$'s outcome depends only on unit $i$'s exposure. While spatial interference is a prevalent and challenging problem in spatial causal inference \citep{khot2025spatial,papadogeorgou2023spatial, reich2021a}, we adopt the standard no-interference assumption here. 

Assumption \ref{b} is the form of the outcome model. Here, $m$ is an arbitrary fixed function, and $X_i,U_i$ may capture any number of measured and unmeasured confounders, as well as exogeneous error, so the outcome model is left completely general. 

\begin{figure}
\centering

    \fontsize{16}{19}\selectfont
    
\begin{tikzpicture}[scale=0.7,transform shape]

  \tikzset{input/.style={}}
  \tikzstyle{pinstyle} = [pin edge={to-,thick,black}]

  \node [input, name=input] {};
  \node [below = 2.75cm of input] (U) {$U$};
    \node [right = 3cm of input, yshift = -1cm] (Ac) {$A_{C}$};
  \node [right = 3cm of input, yshift = 1cm] (Auc) {$A_{UC}$};
  \node[right = 7cm of input] (A) {$A$};
  \node[above = 3cm of input] (X) {$X$};
\node [right = 2.5 cm of A] (Y) {$Y$};

\begin{scope}[->,>=latex]

\draw[<->] (X) -- (U);
\draw[->] (X) -- (Y);
\draw[->] (X) -- (Ac);
\draw[->] (X) -- (Auc);
\draw[->] (U) -- (Ac);
\draw[->, double, double distance=1.5pt] (Auc) -- (A);
\draw[->, double, double distance=1.5pt] (Ac) -- (A);
\draw[->] (A) -- (Y);
\draw[->] (U) -- (Y);

\end{scope}
\end{tikzpicture}

\caption{Causal graph illustrating assumptions. The double-lined arrows from $A_C$ and $A_{UC}$ to $A$ indicate the deterministic relationship $A = A_{UC} + A_C$, rather than a probabilistic one. The bidirectional arrow between $U$ and $X$ represents the possibility that they share an unobserved common cause, implying dependence between their error terms.} 
\label{fig:dag}
\end{figure}

Assumptions \ref{c}--\ref{d} describe the generation of exposure. These assumptions posit that there exist two scalar random variables, $A_{C}$ and $A_{UC}$, that add 
to form the exposure $A$. This assumption can be relaxed by replacing $A = A_{UC} + A_{C}$ with $A = h(A_{UC}, A_{C})$, where $h$ is an unknown monotonic function in its second argument \citep{imbens2009identification}. However, we retain the additive decomposition to maintain consistency with the spatial confounding methods described in Section \ref{framework}. 

Assumption \ref{d}, arguably the most important assumption, requires $A_{UC}$ to be jointly independent of unmeasured confounders and the variable $A_{C}$ conditional on measured confounders $X$. $A_{UC}$ is an instrument, since it is relevant (Assumption \ref{c}), exogeneous (Assumption \ref{d}), and obeys exclusion restriction (Assumption \ref{b}) conditional on measured confounders $X$.

Given Assumptions \ref{a}--\ref{d}, the following results hold:

\begin{proposition}
\label{g}
    $A_i \indep  U_i\mid(X_i, A_{C_i})$.
\end{proposition}
\begin{proposition}
\label{h}
    $Y_i(a) \indep A_i \mid (X_i, A_{C_i})$ for all $a$.
\end{proposition} 

Proposition \ref{g} asserts that $A$ is independent of unmeasured confounders given $A_{C}$ and measured confounders $X$. Proposition \ref{h} asserts that conditional ignorability holds, conditional on measured confounders $X$ and the variable $A_{C}$. Consequently, under appropriate positivity conditions, many causal effects are identifiable; we provide two examples below.

\begin{corollary}[Identification of the exposure response curve]
\label{i}
The exposure response curve is identified as  
$$\E(Y_i(a)) = \E(\E(Y_i\mid A_i = a,  A_{C_i}, X_i))$$ 
for $a \in  \{a': \text{supp}(A_{C}, X) = \text{supp}(A_{C}, X\mid A = a') \}$.
\end{corollary}

\begin{corollary}[Identification of a truncated exposure effect]
\label{j}
The effect of enforcing exposure levels below a predefined cutoff value $c$ is identified as
$$\frac{\E(Y_i(\min(A_i, c))}{\E(Y_i)} =
\frac{\E(\E(Y_i \mid A_i = c, X_i, A_{C_i})) \mid A_i \geq c)\text{pr}(A_i \geq c) + \E(Y_i\mid A_i < c) \text{pr}(A_i < c)}{\E(Y_i)}$$
if $a>c, (a,a_c,x) \in \text{supp}(A, A_C, X) \Rightarrow (c,a_c,x) \in \text{supp}(A,A_C,X)$.
\end{corollary}

We emphasize that the key innovation of these identification results is the incorporation of $A_C$ into the conditioning set. Once determined, $A_C$ is treated as a measured confounder. For further details on identification and proofs of Propositions \ref{g}--\ref{h}, see the Supplementary Material. 

Assumptions \ref{a}--\ref{d} generalize the restrictive underlying assumptions of the spatial confounding methods of Section \ref{framework} (Assumptions \ref{a1}--\ref{e1}) in the following ways. First, the dimensions of $X$ and $U$ can be arbitrary. Second, the outcome model $m$ is no longer required to be a linear function, and instead can accommodate considerable complexity, such as non-linear functions of exposure and effect heterogeneity by measured and unmeasured confounders. For instance, $m$ may include arbitrary interactions between non-linear functions of $A$, $U$, and $X$.
In these two ways, we have relaxed the restrictive assumptions of the spatial confounding methods of Section \ref{framework}. 
However, we impose a stronger condition on the instrument $A_{UC}$ than Assumptions \ref{c1}--\ref{d1}. Instead of demanding $A_{UC}$ to be merely uncorrelated with both $U$ and $A_{C}$, we require $A_{UC}$ to be jointly independent of $(U, A_{C})$ conditional on measured confounders. Caution should be exercised when applying this assumption, especially in environmental epidemiology applications; we discuss this further in Section \ref{discussion}. 

We briefly contrast our identification strategy with other approaches for addressing unmeasured spatial confounding. \cite{gilbert2021causal} require that the unmeasured confounder is a measurable function of spatial coordinates and that the exposure exhibits non-spatial variation. Similarly, distance adjusted propensity score matching is justified by these same assumptions \citep{papadogeorgou2019a}. \cite{schnell2020a} impose strong parametric assumptions on the joint distribution of the unmeasured confounder and exposure. In contrast, our identification relies on an additive decomposition of exposure, where one component is independent of the unmeasured confounder conditional on the measured covariates. Thus, our assumptions focus more on the mechanism that generates exposure than on the behavior of the unmeasured spatial confounder. These two perspectives are closely related, but in practice, one set of assumptions may seem more plausible than the other. Does the practitioner believe that the unmeasured confounders are continuous functions of space, or do they have strong prior knowledge about the exposure generation process, such as the presence of small-scale, localized variation that is independent of the confounders? 

In the following two sections, we focus on estimating a truncated exposure effect, which quantifies the effect of enforcing exposure levels below a predetermined cutoff. This causal effect offers three advantages over the exposure response curve. First, the truncated exposure effect captures a realistic and policy‐relevant intervention that is frequently examined in environmental regulation, for example in studies of air pollution standards \citep{diaz2013assessing, tec2024causal}. Second, its estimation relies on a considerably weaker positivity than that required for identifying the exposure response curve. The latter requires that every unit has positive probability of receiving exposure level $a$, whereas the truncated exposure effect only requires that units with exposure levels above $c$ have positive probability of receiving exposure level $c$. Third, the scalar nature of this causal effect concisely summarizes the outcomes of our framework.

We consider two choices for the instrument $A_{UC}$, both motivated by the premise that small-scale spatial variation in exposure is more likely to be conditionally independent of the unmeasured confounder (Assumption \ref{d}). The first uses a thin plate spline basis to decompose the exposure into small- and large-scale components. Specifically, $A_{UC}$ is constructed as the residuals from a thin plate spline regression of exposure on latitude and longitude, and $A_C$ as the predicted values, following \cite{dupont2022a}. The second uses a Graph Laplacian basis for the decomposition. $A_{UC}$ is constructed as the projection of exposure onto high-frequency eigenvectors of the Graph Laplacian corresponding to large eigenvalues, and $A_C$ as its projection onto the remaining eigenvectors, following \cite{guan2022spectral, urdangarin2024simplified}. For estimation of the truncated exposure effect, now adjusting for $A_{C}$ as a measured confounder, we apply the methodology by \cite{kennedy2017non} using the npcausal package. This approach finds a doubly robust mapping whose conditional expectation given exposure for values exceeding $c$ equals $E(E(Y_i\mid A_i = c, A_{C_i}, X_i)\mid A_i \geq c)$ as long as either the conditional exposure density or outcome model is correctly specified. See the Supplementary Material for additional detail. 

\section{Numerical example}
\label{simulation}

Using the spatial structure of US counties, we create datasets subject to an unmeasured spatial confounder that affects both the exposure and outcome. We estimate the effect of enforcing exposure levels below $c= 0.5$ using our proposed methodology, leveraging localized spatial variation in exposure as an instrument. We investigate the performance of our approach across five confounding mechanisms that differ in the spatial scale and structure of the unmeasured confounding and under varying levels of outcome model complexity. All simulation code can be found at \texttt{https://github.com/NSAPH-Projects/iv-spatialconfounding}. 

We access the U.S. Census Bureau 2010 TIGER/Line Shapefiles to obtain spatial coordinates of county centroids in the contiguous United States \citep{uscensus2010_tiger_counties}. We restrict our simulation to the $n = 503$ counties in the sixth Environmental Protection Agency region (New Mexico, Texas, Oklahoma, Arkansas, and Louisana) to reduce the burden of computation. For $i = 1, \ldots, n = 503$, we generate ($A_{UC}$, $A_{C}$, $U$) using five different mechanisms, varying the structure of the unmeasured spatial confounding. Following \cite{paciorek2010a}, four of the five confounding mechanisms generate ($A_{UC}$, $A_{C}$, $U$) using Gaussian processes with Mat\'ern spatial correlation functions $R( \theta, \nu=2)$, where distance is measured in units of $10^6$m. Two additional confounding mechanisms are considered in the Supplementary Material. 

The first confounding mechanism generates: 
\begin{align*}
    \begin{pmatrix} A_{UC} \\
    A_{C}\\
    U
    \end{pmatrix} &\sim N\bigg\{\begin{pmatrix} (0.1)1_n \\ (-0.2)1_n \\ (0.3)1_n \end{pmatrix}, \begin{pmatrix} R(\theta_{A_{UC}}) & 0 & 0 \\
    0 & R(\theta_{A_{C}}) & 0.95 R(\theta_{A_{C}})\\
    0 & 0.95 R(\theta_{A_{C}}) & R(\theta_{A_{C}})
    \end{pmatrix}\bigg\},
\end{align*} 
\noindent with $\theta_{A_{UC}} = 0.01$ and $\theta_{A_{C}} = 0.5$, so that the spatial range of the unconfounded component of exposure is much smaller than that of the confounded component. The second confounding mechanism generates spatially correlated random fields for $A_C, U$ with a bivariate Leroux conditional autoregressive model. The third confounding mechanism uses the same Gaussian process as the first, applied independently across states. This represents an unmeasured spatial confounder that is continuous within states but discontinuous between them. For example, if $A$ is air pollution, an unmeasured confounder $U$ could be healthcare access, which may vary smoothly within a state but shift abruptly at state borders due to differing policies or funding. The fourth confounding mechanism incorporates two unmeasured confounders. Under the fifth confounding mechanism, the relative spatial scales of $A_{UC}$ and $A_C$ are reversed by setting $\theta_{A_{UC}} = 0.1$ and $\theta_{A_{C}} = 0.01$, so that $A_{UC}$ is smoother than $A_C$. Finally, the exposure $A$ is generated as $A = A_{UC} + A_{C}$ across all confounding mechanisms. By design, each confounding mechanism satisfies the assumptions \ref{a}--\ref{d} of Section \ref{erf}. The Supplementary Material provides detailed descriptions of the confounding mechanisms and plots one realization of $ (A_{UC}, A_{C}, U)^T$ for each.

We further consider two possible outcome models. The linear outcome model generates outcome as $Y_i \sim \mathcal{N}( -0.5 + A_i -U_i - 0.5A_i U_i, 1).$ 
The non-linear outcome model generates outcome as $Y_i \sim \mathcal{N}(-0.5 - 0.5 U_i + \tanh(1.5 A_i) - 0.2 U_i \tanh A_i + 0.1 (\tanh A_i)^2, 1).$ For each of the $10$ data-generating scenarios produced by the five confounding mechanisms and two outcome models, we create $M = 1000$ datasets of size $n = 503$. 

Under each data-generating scenario, we flexibly estimate the truncated exposure effect $E(Y_i(\min(A_i, 0.5)))/ E(Y_i)$ using seven approaches. Each approach employs the same doubly robust estimation procedure \citep{kennedy2017non} but adjusts for a different set of confounders. The oracle approach adjusts for the unmeasured confounder. The baseline approach does not adjust for any confounders. The spatial coordinates adjusts for latitude and longitude, following  \cite{gilbert2021causal}. 

Our four proposed approaches decompose the exposure into small- and large-scale variation, such that the small-scale variation serves as a candidate instrument and the large-scale variation is adjusted for as a measured confounder. Specifically, IV-TPS fits an unpenalized thin plate spline regression of exposure on latitude and longitude with $35$ degrees of freedom, and adjusts for the predicted values from this regression, drawing on \cite{dupont2022a, keller2020a}. IV-GL fits a regression of exposure onto the smoothest $35$ eigenvectors of the Graph Laplacian---corresponding to the $35$ lowest eigenvalues---and adjusts for the predicted values from this regression, drawing on \cite{guan2022spectral, urdangarin2024simplified}.
The choice of dimension $\lfloor 0.07n \rfloor = 35$ is motivated by recommendations from \cite{urdangarin2023evaluating}. IV-TPS+spatialcoord and IV-GL+spatialcoord extend IV-TPS and IV-GL by additionally adjusting for spatial coordinates to improve precision. Under the fifth confounding mechanism, where the relative scales of $A_{UC}$ and $A_C$ are switched, the IV methods instead adjust for the residuals instead of the predicted values.

We evaluate the performance of the seven approaches in estimating the truncated exposure effect through bias and root mean squared error relative to the oracle mean. The four proposed approaches IV-GL, IV-TPS, IV-GL+spatialcoord, and IV-TPS+spatialcoord generally perform very similarly across the ten data-generating scenarios. Each estimates a truncated exposure effect with reduced bias compared to the baseline; in scenarios 1 and 4, the estimates are nearly unbiased. We find that spatial coordinates produces estimates with slightly smaller uncertainty than our proposed methods, but encounters larger bias. We highlight several takeaways. 

\begin{figure}
    \centering
    \includegraphics[width=\linewidth]{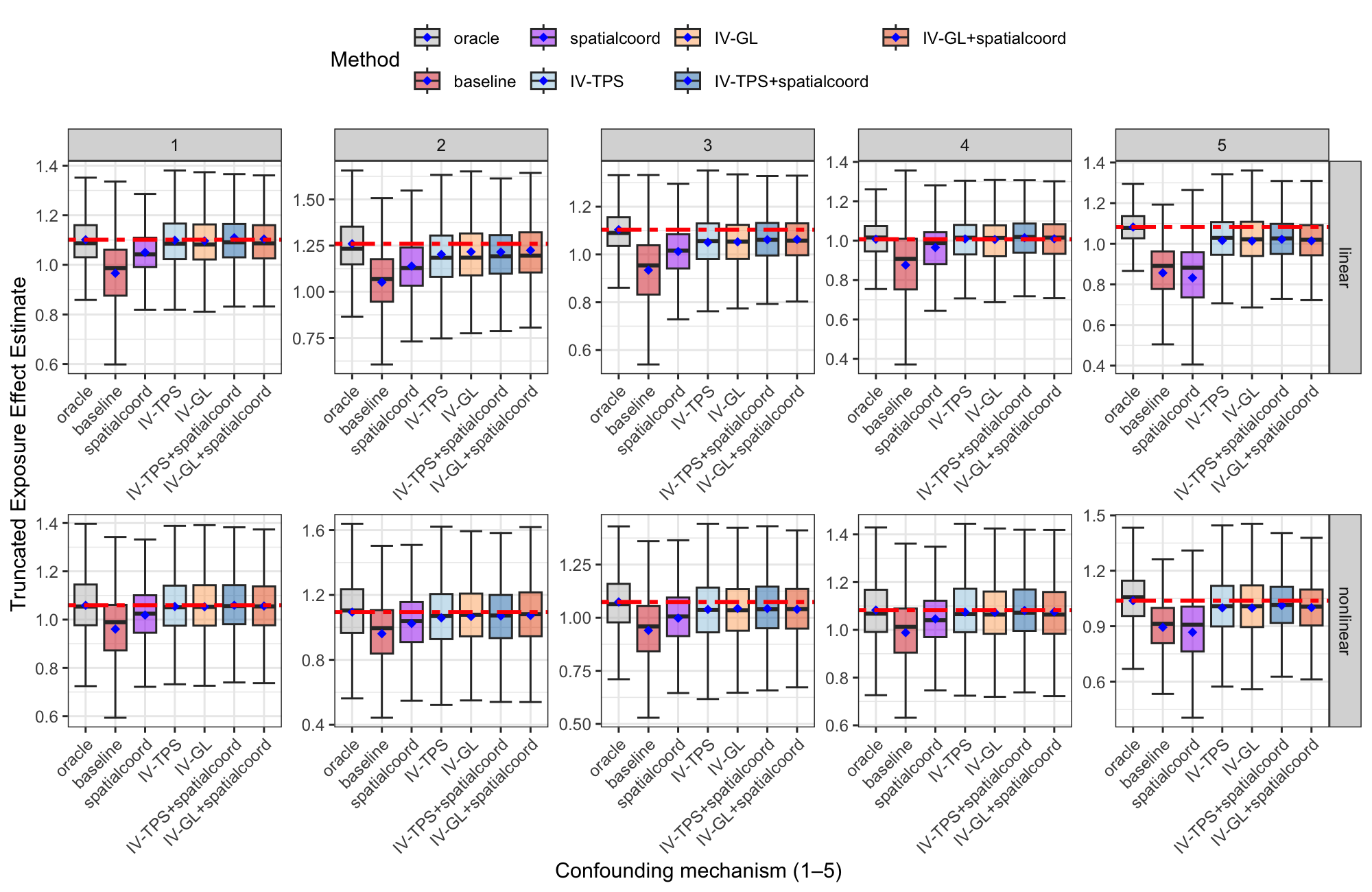}
    \caption{Boxplots of the truncated exposure effect estimates across $1000$ simulations for each combination of outcome model (linear or non-linear) and confounding mechanism (1, 2, 3, 4, or 5). The blue diamonds indicate the means of the estimates, and the dashed red lines correspond to the means of the oracle estimates.}
    \label{fig:combined_tall_results}
\end{figure}

 \begin{enumerate}
 \item The data-generating scenarios with non-linear outcome models differ little from those with linear outcome models, apart from greater variance in the estimates. All seven methods use the same estimation strategy, differing only in the covariates included: flexible non-linear regressions are fit for the treatment density and outcome mean models and then combined into a final estimate. As a result, each method is well equipped to accommodate interaction terms and non-linear functions of exposure and confounder in the outcome model.
     \item The impact of basis mismatch---that is, constructing $A_C$ and $A_{UC}$ with the wrong spatial basis---is not always severe. The validity of IV-TPS and IV-GL hinges on their ability to isolate exposure variation independent of unmeasured confounding; but in practice, the unconfounded component of exposure is unknown. To reflect this challenge, we generated exposures from Gaussian processes and a bivariate Leroux conditional autoregressive model rather than from thin plate spline or Graph Laplacian bases, thereby introducing deliberate basis mismatch. However, both IV-TPS and IV-GL reduced bias across all confounding mechanisms relative to the baseline, even under severe mismatch---particularly in mechanisms 2, 3, and 5, which involved Leroux models, state-specific Gaussian processes, and scale reversal. 
 For comparison, analyses in the Supplementary Material using the true valid IV (trueIV and trueIV+spatialcoord) eliminated all bias.
 \item The IV methods can accommodate multiple unmeasured confounders. This claim is supported by results under the fourth confounding mechanism, where all four IV-based methods effectively remove the bias from two unmeasured confounders. The theory in Section \ref{erf} establishes that identification remains possible, regardless of the number of unmeasured confounders, provided that some unconfounded spatial variation in exposure persists at a distinct spatial scale.
 \item Contrary to common belief, unmeasured confounding bias can be mitigated when the confounded exposure component $A_C$ varies at a smaller spatial scale than the unconfounded component $A_{UC}$ \citep{paciorek2010a}. Our theory indicates that, provided Assumptions \ref{a}--\ref{d} hold for some decomposition of $A$ into $A_C$ and $A_{UC}$, correctly adjusting for measured covariates and $A_C$ can eliminate bias. In confounding scenario 5, where the spatial scales of $A_C$ and $A_{UC}$ are reversed, the IV methods adjust for residuals from the thin plate spline and Graph Laplacian regressions rather than the predicted values. Because these residuals are imperfect proxies for the true $A_C$, some bias persists but is substantially attenuated. However, the methods described in the Supplementary Material that use the true valid IV eliminate the bias entirely.
 \item The spatial coordinates method performs reasonably well even when its identifying conditions are not fully satisfied. Adjusting for spatial coordinates can mitigate unmeasured spatial confounding when the confounder is a measurable, or nearly continuous, function of spatial coordinates and the exposure exhibits non-spatial variation \citep{gilbert2021causal}. These conditions are not satisfied by any of the five confounding mechanisms considered in the main text, yet adjustment for spatial coordinates yields estimates with modest bias and low root mean squared error. 
A mechanism in the Supplementary Material was designed to meet the conditions, but the spatial coordinates method does not consistently outperform the IV-based approaches.
\item Combining IV-TPS or IV-GL with a spatial coordinates adjustment yields nearly unbiased estimates with lower root mean squared errors than IV-TPS or IV-GL alone, highlighting a promising synergy between the two approaches. While each method addresses distinct aspects of unmeasured spatial confounding, their integration appears to enhance robustness by leveraging complementary assumptions about its structure. Further exploration of this hybrid approach may offer new insights into optimizing spatial confounding adjustment.
 \end{enumerate}

\section{Exposure to air pollution and all-cause mortality}

\label{casestudies}

We apply the proposed methodology to estimate the effect of enforcing long-term average air pollution levels below cutoff values of $6$--$12 \mu g/m^3$ on  
all-cause mortality across zip codes in the contiguous United States. As in \citet{tec2024spatial}, we aim to determine whether our approach can effectively adjust for unmeasured spatial confounding by intentionally excluding important spatially structured confounders and verify whether the estimation can recover the original truncated exposure effect estimate.

The exposure $A$ is average fine particulate matter (PM$_{2.5}$) over the period 2001--2010 estimated at the 1km $\times$ 1km grid-level \citep{di2019ensemble}, and the outcome $Y$ is all-cause mortality rate among 68.5 million Medicare enrollees ($\geq 65$ years of age) over the period 2011--2016. Both the exposure and outcome are aggregated to the zip code-level ($n = 33,255$). We additionally consider $14$ zip code-level covariates measured in 2000, including sociodemographic variables collected from the U.S. Census, American Community Survey, and the CDC's Behavioral Risk Factor Surveillance System, as well as four meteorological variables from Gridmet via Google Earth Engine. For additional details on the data and data sources, see the Supplementary Material. 
  
   We estimate the truncated exposure effects $\E(Y(\min(A, c)))/\E(Y)$ for $c \in \{6,\ldots,12\}$ using seven different confounding adjustments within the doubly robust estimation method by \citep{kennedy2017non}. The first approach adjusts for all $14$ measured confounders and is referred to as the oracle. Assuming consistency, positivity, and ignorability (conditional on these $14$ confounders), and provided that the conditional exposure density or the outcome model is correctly specified, this estimate of the truncated exposure effect would be consistent for the true truncated exposure effect subject to additional regularity conditions. The second approach excludes four temperature and humidity variables,  
   but adjusts for the remaining $10$ measured covariates. We refer to this approach as the baseline, as it represents an estimate of the 
  truncated exposure effect subject to unmeasured spatial confounding bias. The third approach adjusts for the remaining $10$ measured covariates as well as spatial coordinates. 
   
\begin{figure}
       \centering
\includegraphics[width=0.8\linewidth]{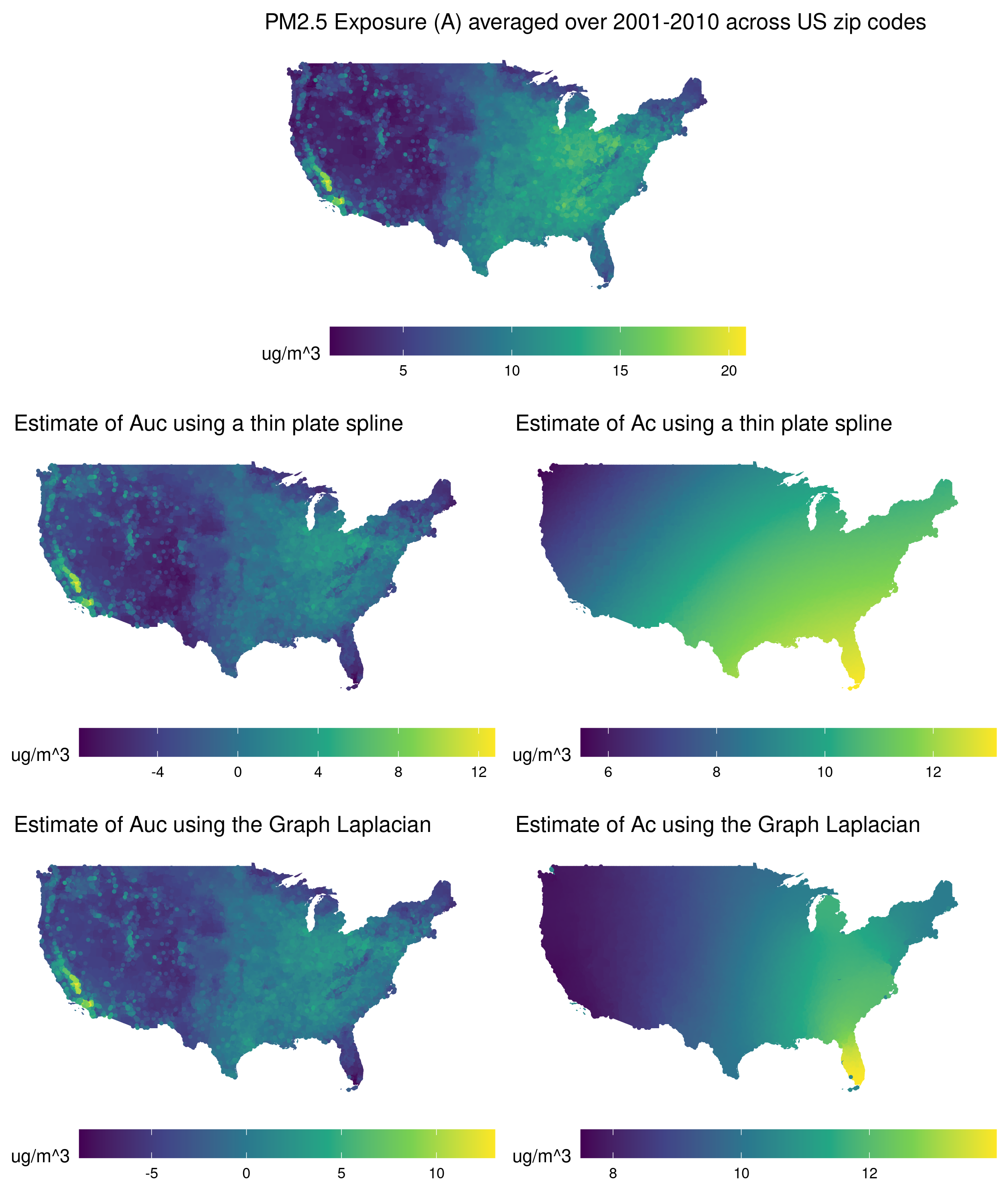}\\
       \caption{Long-term average exposure to PM$_{2.5}$ at the zip code level during 2001--2010, two candidate IVs, and two candidate adjustment variables. Both Post Office boxes (represented with point shapefiles) and Zip Code Tabulation Areas (represented with polygon shapefiles) are plotted above. Shapefiles sourced from ESRI, 2010.}
       \label{fig:exposure_ivs}
   \end{figure}

   The fourth and fifth approaches, IV-TPS and IV-GL, implement our proposed methodology from Section \ref{erf}. Both approaches extract small-scale spatial variation in air pollution as the instrument $A_{UC}$ and adjust for the remaining large-scale spatial variation $A_{C}$ alongside the remaining $10$ measured covariates. Due to the substantial computational burden of calculating basis elements, and given the spatially smooth nature of air pollution exposure, we chose to use bases that captured approximately 20\% of the variance in exposure. The Supplementary Material assesses the sensitivity of the results to the choice of basis dimension. The sixth and seventh approaches, IV-TPS+spatialcoord and IV-GL+spatialcoord, extend the fourth and fifth approaches by additionally adjusting for spatial coordinates. Figure \ref{fig:exposure_ivs} presents the exposure, the two candidate instrumental variables $A_{UC}$, and the two candidate adjustment variables $A_{C}$. 

Previous studies suggest that sharp spatial patterns in exposure to air pollution result from random fluctuations in wind patterns or wildfire smoke, and are therefore independent of unmeasured confounders \citep{bondy2020crime, cabral2024air, gu2020air,jayachandran2009air, schwartz2017estimating, schwartz2018national, 
 yang2018air}. We adopt a similar assumption, hypothesizing that our two candidate instruments $A_{UC}$, as shown in the left panels of Fig. \ref{fig:exposure_ivs}, represent localized spatial variation in air pollution exposure and are independent of the omitted temperature and humidity variables, conditional on the remaining $10$ measured confounders. 

\begin{figure}
        \centering
        
        \includegraphics[width = \linewidth]{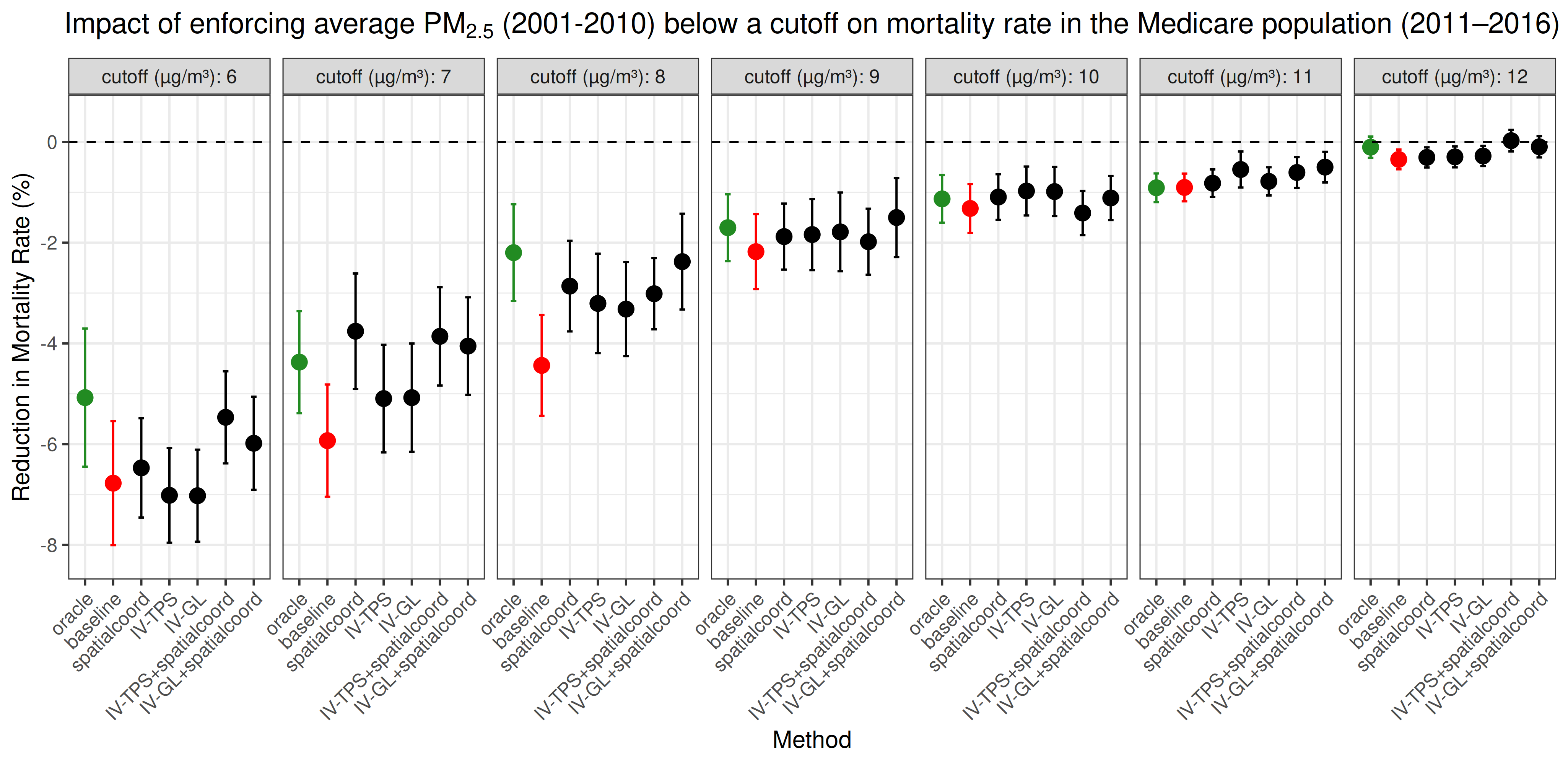}
        \caption{Estimated effect of enforcing average PM$_{2.5}$ below a cutoff value of $6,7,8,9,10,11,$ and $12 \mu g/m^3$ respectively during 2001--2010 on the all-cause mortality rate during 2011--2016 among Medicare enrollees using seven different confounding adjustments. The y-axis is $100\%[E(Y(\min(A, c)))/\E(Y)-1]$.} 
        \label{fig:erfs}
    \end{figure}

   Figure \ref{fig:erfs} presents the truncated exposure effect estimates from each of the seven methods, along with corresponding confidence intervals. Details on uncertainty quantification are provided in the Supplementary Material. For the cutoff value $6 \mu g/m^3$, effect estimates range from $0.93$ to $0.95$, suggesting a significant beneficial effect of reducing air pollution levels to below the standard. For the cutoff value $12 \mu g/m^3$, effect estimates range from $0.997$ to $1$ with smaller uncertainty. The oracle estimate exceeds the baseline estimate for all cutoff values, suggesting the presence of unmeasured spatial confounding due to the omission of temperature and humidity variables. 
   
  We emphasize two key findings. First, the bias resulting from the unmeasured spatial confounding due to the exclusion of the temperature and humidity variables is reasonably small for all cutoff values. This finding is consistent with those of \cite{wu2020evaluating}, whose sensitivity analyses indicated that point estimates do not vary much when temperature and humidity were excluded. Second, the estimates of the spatial coordinates method and four proposed methods are generally higher than the baseline estimate, demonstrating their ability to attenuate the confounding bias. However, this pattern is not consistent across cutoff values, indicating that the necessary spatial confounding adjustment may depend on the estimand.
  
  IV-GL+spatialcoord produced estimates and confidence intervals closest to the oracle, as measured by the average Hausdorff distance across the seven cutoff values, followed by IV-TPS+spatialcoord. For further details on this metric see the Supplementary Material. As in Section \ref{simulation}, these findings further underscore the promising synergy between our IV methods and the spatial coordinates approach proposed by \cite{gilbert2021causal}. Nonetheless, we caution against over-interpreting these results, as the oracle itself may be subject to residual unmeasured confounding and may not represent an unbiased estimate of the true truncated exposure effect.  
    
    Despite these differences, all truncated exposure effect estimates suggest that reducing PM$_{2.5}$ exposure below the current National Ambient Air Quality Standard of $9\mu g/m^3$ during 2001–2010 would have significantly lowered the average all-cause mortality rate among Medicare enrollees from 2011 to 2016 by $1$--$3\%$ across U.S. zip codes. This effect diminishes in magnitude as the cutoff increases, indicating that strengthening air pollution standards would have significantly reduced mortality during this time period. Our results are reasonably consistent with the effect sizes reported in other studies of air pollution and mortality \citep{beelen2014effects,di2017air,dockery1993association,liu2019ambient, pappin2019examining, wu2020evaluating}.  
    \section{Discussion}
\label{discussion}
In this paper, we introduce a novel IV framework that redefines many of the existing approaches to adjust for unmeasured confounding as IV methods. This new perspective represents a paradigm shift in the study of spatial confounding. By reinterpreting standard spatial confounding methods through the lens of 
IVs, our framework unifies existing approaches by showing that many are built on a common theoretical foundation: they assume that the exposure can be decomposed additively into two components, a large-scale component correlated with the unmeasured confounder and an uncorrelated small-scale component. The small-scale component is then leveraged as an instrumental variable to estimate the parameter of interest. While prior work by \cite{giffin2021instrumental} has proposed using instrumental variables to address spatial confounding and interference, the broader link between spatial confounding methods and the instrumental variables literature has not yet been fully established. By placing spatial confounding methods within a unified framework, we bring coherence to a previously fragmented literature and show that the methods differ primarily in the spatial basis used to decompose the exposure.
This perspective also provides a foundation for extending these approaches to estimate a broader class of causal effects, including those that do not require linearity in the outcome model. The results of the simulation and data application suggest a promising synergy between our IV methods and the spatial coordinates approach proposed by \cite{gilbert2021causal}.

Our identification approach is closely related to recent work on network interference \citep{gao2024endogenous, li2022random, lu2024estimation}, where consistent estimation is achieved by projecting out signal components aligned with the graph’s leading eigenvectors. It also shares similarities with the front-door criterion \citep{pearl1995causal, pearl2009causality}, which enables identification through downstream unconfounded mediators. Our approach is the mirror image; identification relies on an upstream, unconfounded component of exposure. More broadly, these parallels illustrate a common identification strategy: positing that the exposure arises from, or gives rise to, multiple sources of variation, with some components potentially confounded and others available as unconfounded variation for identification.

This decomposition-based view of identification suggests several applications beyond spatial confounding. 
For temporal exposures such as temperature, long-term trends tend to co-vary with seasonal and meteorological factors, whereas short-term fluctuations are more likely to vary independently of unmeasured confounders \citep{dominici2003airborne}. 
Here, $A_{UC}$ and $A_C$ could be obtained by decomposing the exposure into short- and long-term variation.
For hierarchical data, between-group differences in exposure may reflect unobserved contextual factors common to individuals of the same group, 
while within-group differences may be less influenced by such unmeasured confounding \citep{diez2000multilevel}. Here, $A_{UC}$ and $A_C$ could be obtained from a within- and between-group decomposition. 
For mixture exposures, individual constituents such as air pollutants or chemicals often share upstream sources that may be correlated with unmeasured confounders like industrial activity or land use \citep{braun2016can}. Here, $A_{C}$ and $A_{UC}$ could be obtained by decomposing the mixture into variation arising from shared sources and residual variation.

There are several opportunities for future research. First, while our work proposed a fundamental framework to unify existing methods and distinguish the type and dimension of basis used to construct $A_{UC}$ and $A_{C}$, specifying these choices for spatial confounding adjustment underscores the need for a sensitivity analysis framework. The Supplementary Material provides a sensitivity analysis for the basis dimension used in the two spatial decompositions. In theory, dimension of the basis induces a trade-off between bias and variance: increasing the dimension removes large-scale spatial variation in exposure, potentially isolating unconfounded variation and producing unbiased causal estimates, but at the cost of increased variance \citep{dominici2004improved}. Our sensitivity analysis shows that this trend emerges when measured covariates are omitted, while estimates remain reasonably stable when they are included. A potential area for investigation is developing a selection procedure for the basis dimension, akin to \cite{keller2020a}, but without requiring parametric assumptions and for more general causal effects. 

Another possible direction, to avoid making assumptions about the spatial basis of the unmeasured confounding altogether, is to extend the ``basis voting'' method of \cite{burman2025robust} to more flexible, nonparametric outcome model settings. 
Under suitable conditions, one might identify multiple candidate decompositions of the exposure into confounded and unconfounded components, obtain several consistent causal effect estimates, and combine them to form a more efficient estimator.

A second open question concerns the existence of $A_{UC}$, that is, whether any variation in exposure is unconfounded.
In such cases, sensitivity parameters could be used to quantify deviations from the unconfoundedness assumption and establish point or set identification of the estimand as a function of these parameters \citep{ding2016sensitivity, rosenbaum1983assessing}. Incorporating spatiotemporal data or an auxiliary dataset may provide an additional avenue for verifying or identifying instrumental variation. 

\subsection*{Acknowledgments}
The authors would like to thank Heejun Shin and James Kitch for their thoughtful comments. The computations in this paper were run on the FASRC Cannon cluster supported by the FAS Division of Science Research Computing Group at Harvard University. 

\subsection*{Funding}
The authors gratefully acknowledge support from the National Institutes of Health under award numbers R01ES030616, R01AG066793, RF1AG074372-01A1, R01MD016054, R01ES034373, RF1AG080948, U24ES035309, RF1AG071024, P30ES000002, R01ES34021, R01ES037156-01, T32ES007142; the Sloan Foundation G-2020-13946; and the National Science Foundation Graduate Research Fellowship under Grant No. DGE 2140743. Any opinion, findings, and conclusions or recommendations expressed in this material are those of the authors and do not necessarily reflect the views of our funders. 

\bibliographystyle{apalike} 
\bibliography{main}      

\newpage
\singlespacing

\tableofcontents
\doublespacing

\makeatletter
\renewcommand \thesection{S\@arabic\c@section}
\renewcommand\thetable{S\@arabic\c@table}
\renewcommand \thefigure{S\@arabic\c@figure}

\renewcommand{\arraystretch}{0.7}
\makeatother
\section{IV framework unifying six spatial confounding methods}
Below, we demonstrate that the six methods for addressing spatial confounding \citep{dupont2022a, urdangarin2023evaluating, keller2020a, guan2022spectral, thaden2018structural, wiecha2024two} are particular instances of the general framework we propose.

\subsection{2SLS methods}
 \cite{dupont2022a} and \cite{ urdangarin2024simplified} apply two-stage least squares (2SLS) using $A_{UC}$ as an instrument. In the first stage, exposure is decomposed into large-scale spatial variation ($A_C$) and small-scale spatial variation $(A_{UC})$. In the second stage, exposure is replaced with $A_{UC}$ in the outcome regression \citep{baiocchi2014instrumental, terza2008two, greene2003econometric}. 

Our framework formally establishes assumptions that guarantee the validity of this approach. Specifically, if Assumptions 1–5 hold, then the coefficient estimate of $A_{UC}$ from the second-stage regression converges in probability to \begin{align*}
    \frac{\text{Cov}(Y, A_{UC})}{\text{Var}(A_{UC})} &= \frac{\text{Cov}(\beta_0 + \beta A + \epsilon, A_{UC})}{\text{Var}(A_{UC})} = \frac{\text{Cov}(\beta A_C + \beta A_{UC} + \epsilon, A_{UC})}{\text{Var}(A_{UC})} = \beta,
\end{align*}the statistical parameter of interest. The first equality follows from Assumption 1, the second from Assumption 2, and the third from Assumptions 3-5.

We now describe the specific forms of $A_C$ and $A_{UC}$ as used in \cite{dupont2022a, urdangarin2024simplified}.

\begin{enumerate}
    \item The spatial+ method proposed by \cite{dupont2022a}, in its unsmoothed form, is a 2SLS method 
that decomposes exposure into large-scale and small-scale spatial variation using a \textbf{thin plate spline basis}. The assumed data-generating process is 
\begin{align*}
    A_i &= g(S_i) + \epsilon_i^A,\\
    Y_i &= \beta_0 + \beta A_i + f(S_i) + \epsilon_i^Y,
\end{align*}for $i = 1, \ldots, n$, where $f,g$ are unknown, bounded functions, $S$ denotes spatial coordinates, and $\epsilon_i^A \stackrel{i.i.d}{\sim}\mathcal{N}(0,\sigma^2_A)$, $\epsilon_i^Y \stackrel{i.i.d}{\sim}\mathcal{N}(0,\sigma^2_Y)$.

The first stage regression fits a thin plate spline of spatial coordinates to exposure, obtaining fitted values $\hat{g}(S)$ and residuals $A-\hat{g}(S)$. In the second stage, the outcome is regressed on the first-stage residuals $A-\hat{g}(S)$ and a thin plate spline of spatial coordinates $h$ with the same degrees of freedom as the first stage:
$$Y_i = \beta_0 + \beta(A_i-\hat{g}(S_i))  + h(S_i) + \epsilon_i.$$

Spatial+ falls within our framework by recognizing that the first-stage residuals $A - \hat{g}(S)$ correspond to the instrument $A_{UC}$, as summarized in the table below: 

\begin{center}
    \begin{tabular}{r|l}
    Instrument ($A_{UC}$) & $A-\hat{g}(S)$\\
    Large-scale spatial variation ($A_C$) & $\hat{g}(S)$\\
    Assumption 1 & $ Y_i = \beta_0 + \beta A_i + f(S_i) + \epsilon_i^Y$\\
    Assumption 2 & $A_i = (A-\hat{g}(S)) + (\hat{g}(S))$\\
    Assumption 3 & $ (A-\hat{g}(S)) \perp \hat{g}(S)$\\
    Assumption 4 & $(A-\hat{g}(S)) \perp (f(S) + \epsilon_i^Y)$\\ 
    Assumption 5 & $\text{Var}(A - \hat{g}(S))>0 $
\end{tabular}
\end{center}

If Assumptions 1--5 are satisfied, the second stage regression of spatial+ yields a consistent estimate of $\beta$.

\item The simplified spatial+ method proposed by \cite{urdangarin2024simplified} is a 2SLS method that decomposes exposure into large-scale and small-scale spatial variation using the \textbf{eigenvector basis of a spatial precision matrix}. The authors state that eigenvectors of a spatial precision matrix capture spatial information at different spatial scales, with those corresponding to the lowest non-zero eigenvalues representing the smoothest spatial patterns. The assumed data-generating process is
\begin{align*}
    Y_i | R_i &\sim \text{Poisson}(e_i R_i),\\
    \log R_i &= \alpha + \beta A_i + \theta_i, 
\end{align*}for $i = 1, \ldots, n$, where $e_i$ is expected counts for unit $i$ and $\theta_i$ is a spatial random effect. In the first stage, exposure $A$ is decomposed into large-scale and small-scale spatial variation using the eigenvectors of the spatial precision matrix $\Omega$ for the random effects $(\theta_1, \ldots, \theta_n)$. The large-scale component is obtained by projecting $A$ onto the subspace spanned by the $k+1$ eigenvectors $v_{n-k}, \ldots, v_n$ of $\Omega$ corresponding to the smallest $k+1$ eigenvalues $\lambda_{n-k}, \ldots, \lambda_n$:
$$\sum_{i = n-k}^n v_iv_i^TA.$$
The small-scale component is obtained by projecting $A$ onto the subspace spanned by the remaining $n-(k+1)$ eigenvectors $v_{1}, \ldots, v_{n-(k+1)}$, corresponding to the highest $n-(k+1)$ eigenvalues $\lambda_{1}, \ldots, \lambda_{n-(k+1)}$:
$$\sum_{i = 1}^{n-(k+1)} v_iv_i^TA.$$

In the second stage, the regression model is fit while replacing the exposure $A$ with its small-scale variation $\sum_{i = 1}^{n-(k+1)} v_iv_i^TA$, 
$$\log R = 1_n \alpha + \bigg(\sum_{i = 1}^{n-(k+1)} v_iv_i^TA\bigg)\beta + \theta.$$

Simplified spatial+ falls within our framework by recognizing that the small-scale spatial variation $\sum_{i = 1}^{n-(k+1)} v_iv_i^TA$ corresponds to the instrument $A_{UC}$, as summarized in the table below: 
\begin{center}
    \begin{tabular}{r|l}
    Instrument ($A_{UC}$) & $\sum_{i = 1}^{n-(k+1)}  v_iv_i^TA$\\
    Large-scale spatial variation ($A_C$) & $\sum_{i = n-k}^n v_iv_i^TA$\\
    Assumption 1 & $ \log R_i = \alpha + \beta A_i + \theta_i$\\
    Assumption 2 & $A = (\sum_{i = 1}^{n-(k+1)} v_iv_i^TA) + (\sum_{i = n-k}^n  v_iv_i^TA) $\\
    Assumption 3 & $\sum_{i = 1}^{n-(k+1)}  v_iv_i^TA \perp \sum_{i = n-k}^n  v_iv_i^TA$\\
    Assumption 4 & $\sum_{i = 1}^{n-(k+1)} v_iv_i^TA \perp \theta$\\ 
    Assumption 5 & $\sum_{i = 1}^{n-(k+1)} v_iv_i^TA$ nonconstant
\end{tabular}
\end{center}
If Assumptions 1--5 are satisfied, the second stage regression of simplified spatial+ yields a consistent estimate of $\beta$.

\end{enumerate}

\subsection{2SRI methods}
\cite{guan2022spectral} and \cite{keller2020a} use $A_{UC}$ as an instrument in two-stage residual inclusion (2SRI). The first stage of 2SRI is identical to the first stage of 2SLS. In the second stage, exposure is replaced with $A_{UC}$ and $A_C$ is included as an additional regressor \citep{terza2008two, hausman1978specification}. 

Our framework formally establishes assumptions that guarantee the validity of this approach. Specifically, if Assumptions 1--5 hold, then the coefficient estimate of $A_{UC}$ from the second-stage regression converges in probability to \begin{align*}
    \frac{\text{Var}(A_C)\text{Cov}(Y, A_{UC}) - \text{Cov}(A_{UC}, A_C)\text{Cov}(Y, A_C) }{\text{Var}(A_C)\text{Var}(A_{UC}) - (\text{Cov}(A_{UC}, A_C))^2 )} &= \frac{\text{Cov}(Y, A_{UC})}{\text{Var}(A_{UC})} = \beta,
\end{align*}where the first equality follows by Assumption 3 and the second from Assumptions 1--5 following the 2SLS case.  

\begin{enumerate}
    \item The ``preadjustment of exposure" method proposed by \cite{keller2020a} (Section 2.3) is a 2SRI method that suggests decomposing exposure into large-scale and small-scale spatial variation using any type of \textbf{hierarchical spatial basis} $H$. A hierarchical spatial basis is a basis whose elements are ordered by some notion of spatial scale, such as a thin plate spline basis, Fourier basis, or wavelet basis. The assumed data-generating process is \begin{align*}
        Y_i &= \beta_0 + \beta A_i + f(S_i) + \epsilon_i, 
    \end{align*}for $i = 1, \ldots n$.
    In the first stage, exposure $A$ is decomposed as $$A = H_m(H_m^T H_m)^{-1}H_m^T A + (A -H_m(H_m^T H_m)^{-1}H_m^T A), $$
    where $H_m$ consists of the smoothest or largest-scale basis vectors $m$ of the basis $H$.

    In the second stage, the regression model is fit while replacing the exposure $A$ with its small-scale variation and including large-scale spatial variation as a covariate,
    $$Y = \beta_01_n + \beta(A-H_m(H_m^T H_m)^{-1}H_m^T A) + \gamma(H_m(H_m^T H_m)^{-1}H_m^T A ) + \epsilon.$$

    The preadjustment of exposure method falls within our framework by recognizing that the small-scale spatial variation $A -H_m(H_m^T H_m)^{-1}H_m^T A$ corresponds to the instrument $A_{UC}$, as summarized in the table below: 

    \begin{center}
    \begin{tabular}{r|l}
    Instrument ($A_{UC}$) & $A -H_m(H_m^T H_m)^{-1}H_m^T A$\\
    Large-scale spatial variation ($A_C$) & $H_m(H_m^T H_m)^{-1}H_m^T A$\\
    Assumption 1 & $ Y_i = \beta_0 + \beta A_i + f(S_i) + \epsilon_i$\\
    Assumption 2 & $A =  (A -H_m(H_m^T H_m)^{-1}H_m^T A) + H_m(H_m^T H_m)^{-1}H_m^T A$\\
    Assumption 3 & $(A -H_m(H_m^T H_m)^{-1}H_m^T A) \perp H_m(H_m^T H_m)^{-1}H_m^T A$\\
    Assumption 4 & $(A -H_m(H_m^T H_m)^{-1}H_m^T A) \perp f(S)$\\ 
    Assumption 5 & $A -H_m(H_m^T H_m)^{-1}H_m^T A$ nonconstant
\end{tabular}
\end{center}
If Assumptions 1--5 are satisfied, the second stage regression yields a consistent estimate of $\beta$.

\item The discrete-space methodology proposed by \cite{guan2022spectral} is a 2SRI method that leverages small-scale variation in exposure as the instrument using the \textbf{eigenvector basis of the Graph Laplacian}.
The eigenvectors $v_1, \ldots, v_n$ of the Graph Laplacian are ordered by a notion of spatial scale \citep{ortega2018graph}. The assumed data-generating process is
\begin{align*}
    Y_i &= \beta_0 + \beta A_i + \gamma U_i + \epsilon_i,
\end{align*}for $i = 1, \ldots n$, where $\epsilon_i \stackrel{i.i.d}{\sim}\mathcal{N}(0,\sigma^2)$ and $U$ is the unmeasured confounder. In the first stage, exposure is decomposed as 
$$A = \sum_{i = 1}^{n-1} v_i v_i^T A +  v_n v_n^T A$$
where the $n$th eigenvector $v_n$ is the smallest-scale, corresponding to the highest eigenvalue $\lambda_n$.

In the second stage, outcome is regressed on the variation of exposure at each spatial scale, $v_1 v_1^T A, \ldots, v_n v_n^T A$:
$$Y = \beta_0 1_n + \sum_{i = 1}^n v_i v_i^T A \bigg(\sum_{l = 1}^L b_l B_l(\omega_i)\bigg) + V + \epsilon,$$
where $B_l(\omega)$ are B-spline basis functions with associated coefficients $b_l$, $V$ denotes spatial random effects from a conditional autoregressive prior, and $\epsilon = (\epsilon_1, \ldots, \epsilon_n)^T$ is i.i.d Gaussian error. 

The posterior of $\sum_{l = 1}^L B_l(\omega_n) b_l$ is used as an estimate of $\beta$; note that this is the coefficient of $v_n v_n^T A$ in the model above.

\cite{guan2022spectral} falls within our framework by recognizing that the small-scale spatial variation $ v_n v_n^T A$ corresponds to the instrument $A_{UC}$, as summarized in the table below: 

\begin{center}
    \begin{tabular}{r|l}
    Instrument ($A_{UC}$) & $ v_n v_n^T A$\\
    Large-scale spatial variation ($A_C$) & $\sum_{i = 1}^{n-1} v_i v_i^T A$\\
    Assumption 1 & $Y = \beta_01_n + \beta A + \gamma U + \epsilon$\\
    Assumption 2 & $A =  v_n v_n^T A + \sum_{i = 1}^{n-1} v_i v_i^T A $\\
    Assumption 3 & $v_n v_n^T A \perp \sum_{i = 1}^{n-1}  v_i v_i^T A $\\
    Assumption 4 & $ v_n v_n^T A \perp U$\\ 
    Assumption 5 & $v_n v_n^T A$ nonconstant
\end{tabular}

\end{center}
If Assumptions 1--5 are satisfied, the second stage regression yields a consistent estimate of $\beta$.

\end{enumerate}

\subsection{Double prediction methods}

\cite{thaden2018structural, wiecha2024two} 
employ $A_{UC}$ as an instrument in double prediction. In the first stage of double prediction, both the exposure and outcome are decomposed into small-scale and large-scale spatial variation, obtaining $A_{UC}$, $A_C$ and $Y_{UC}, Y_C$ respectively.  
In the second stage, the small-scale spatial variation in outcome $Y_{UC}$ is regressed on the small-scale spatial variation in exposure $A_{UC}$.

Our framework formally establishes assumptions that guarantee the validity of this approach. Here, we require the instrument to satisfy an additional assumption:\begin{align*}
    \text{Assumption 6}: A_{UC}\perp Y_C.
\end{align*}Under Assumptions 1--6, the coefficient estimate of $A_{UC}$ from the second-stage regression converges in probability to
$$\frac{\text{Cov}(A_{UC}, Y_{UC})}{\text{Var}(A_{UC})} = \frac{\text{Cov}(A_{UC}, Y - Y_C)}{\text{Var}(A_{UC})} = \frac{\text{Cov}(A_{UC}, Y)}{\text{Var}(A_{UC})} = \beta. $$

\begin{enumerate}
    \item The geoadditive structural equation model (gSEM) proposed by \cite{thaden2018structural} is a double prediction method that leverages small-scale variation in exposure as the instrument using a basis consisting of $d$ \textbf{region-level indicators}. The assumed data-generating process is \begin{align*}
        Y_i &= \beta_0 + \beta A_i + U_i + \epsilon_i,
    \end{align*}where $U_i$ is an unmeasured confounder that is constant within each of the $d$ spatial regions. 
    In the first stage, both exposure and outcome are regressed on the region indicators, obtaining the decompositions
    \begin{align*}
        A_i &= \sum_{k = 1}^d z_{ki} \gamma_{1k} + (A_i - \sum_{k = 1}^d z_{ki} \gamma_{1k}),\\
        Y_i &= \sum_{k = 1}^d z_{ki} \gamma_{2k} + (Y_i - \sum_{k = 1}^d z_{ki} \gamma_{2k}),
    \end{align*} where $z_{ki} \in \{0,1\}$ is the indicator that unit $i$ is located in region $k$ for $k = 1, \ldots, d$, and $\gamma_{1k} = (z_k^t z_k)^{-1} z_k^t A$, $\gamma_{2k} =  (z_k^t z_k)^{-1} z_k^t Y$ if penalization is not used.

    In the second stage, the residuals from the outcome-indicator regression are regressed on the residuals from the exposure-indicator regression:
    $$Y_i - \sum_{k = 1}^d z_{ki} \gamma_{2k} = \beta(A_i - \sum_{k = 1}^d z_{ki} \gamma_{1k}) + \epsilon_i.$$
    gSEM falls within our framework by recognizing that the small-scale variation $A_i - \sum_{k = 1}^d z_{ki} \gamma_{1k}$ corresponds to the instrument $A_{UC}$, as summarized in the table below. 

    \begin{center}
    \begin{tabular}{r|l}
    Instrument ($A_{UC}$) & $A_i - \sum_{k = 1}^d z_{ki} \gamma_{1k}$\\
    Large-scale spatial variation ($A_C$) & $ \sum_{k = 1}^d z_{ki} \gamma_{1k}$\\
    $Y_{UC}$ & $Y_i - \sum_{k = 1}^d z_{ki} \gamma_{2k}$\\
    $Y_C$ & $\sum_{k = 1}^d z_{ki} \gamma_{2k}$\\
    Assumption 1 & $Y_i = \beta_0 + \beta A_i + U_i + \epsilon_i$\\
    Assumption 2 & $A_i = (A_i - \sum_{k = 1}^d z_{ki} \gamma_{1k}) + \sum_{k = 1}^d z_{ki} \gamma_{1k}$\\
    Assumption 3 & $(A - \sum_{k = 1}^d z_{k} \gamma_{1k}) \perp z_{k} \gamma_{1k}$\\
    Assumption 4 & $(A - \sum_{k = 1}^d z_{k} \gamma_{1k})  \perp (U + \epsilon)$\\ 
    Assumption 5 & $(A - \sum_{k = 1}^d z_{k} \gamma_{1k}) \perp \sum_{k = 1}^d z_{k} \gamma_{2k}$\\
    Assumption 5 & $A - \sum_{k = 1}^d z_{k} \gamma_{1k}$ nonconstant
\end{tabular}

\end{center}
If Assumptions 1--6 are satisfied, the second stage regression yields a consistent estimate of $\beta$.

\item Double spatial regression (DSR) proposed by \cite{wiecha2024two} is a double prediction method that uses the residuals from \textbf{Gaussian process regression with Mat\'ern correlation} as an instrument. The assumed data-generating process is \begin{align*}
    Y_i &= \beta A_i + f(S_i) + U_i,\\
    A_i &= g(S_i) + V_i
\end{align*}for $i = 1, \ldots, n$, where $S$ denotes spatial coordinates, $U_i$ and $V_i$ are error terms with finite, non-zero variance such that $\E(U_i|A_i, S_i) = 0, \E(V_i|S_i) = 0$. 

In the first stage, both exposure and outcome are decomposed into large-scale and small-scale spatial variation using kriging: \begin{align*}
    A &= (A - \hat{g}(S)) +\hat{g}(S),\\
    Y &= (Y - \hat{h}(S)) + \hat{h}(S),
\end{align*}where $\hat{g}(S), \hat{h}(S)$ are universal kriging estimates of the spatial trends in exposure and outcome respectively. In the second stage, the kriging residuals are combined to form an estimate of $\beta$:
$$\hat{\beta} = \bigg((A - \hat{g}(S))^T(A - \hat{g}(S))\bigg)^{-1}(A - \hat{g}(S))^T (Y - \hat{h}(S)).$$

Double spatial regression falls within our framework by recognizing that the small scale variation $A - \hat{g}(S)$ corresponds to the instrument $A_{UC}$, as summarized in the table below. 

\begin{center}
    \begin{tabular}{r|l}
    Instrument ($A_{UC}$) & $A - \hat{g}(S)$\\
    Large-scale spatial variation ($A_C$) & $ \hat{g}(S)$\\
    $Y_{UC}$ & $(Y - \hat{h}(S))$\\
    $Y_C$ & $\hat{h}(S)$\\
    Assumption 1 & $Y_i = \beta A_i + f(S_i) + U_i$\\
    Assumption 2 & $A = (A - \hat{g}(S)) + \hat{g}(S)$\\
    Assumption 3 & $(A - \hat{g}(S)) \perp \hat{g}(S)$\\
    Assumption 4 & $(A - \hat{g}(S)) \perp (f(S) + U)$\\ 
    Assumption 5 & $(A - \hat{g}(S)) \perp \hat{h}(S)$\\
    Assumption 5 & $A - \hat{g}(S)$ nonconstant
\end{tabular}
\end{center} 
If Assumptions 1--6 are satisfied, the second stage regression yields a consistent estimate of $\beta$. In practice, however, kriging implicitly involves a bias–variance trade-off that may result in slight deviations from these assumptions.

\end{enumerate}

The following page presents an expanded version of Table 1 from the main text.

\begin{landscape}

\begin{table}
\footnotesize
    \begin{tabular}{p{2.5cm} ||p{3.4cm}
    p{3.2cm} 
    p{3.75cm} 
    p{0.9cm} 
    p{0.6cm}
    p{4.8cm}
    }
    {Paper} &  \makecell[l]{ Spatial basis or method \\ used to obtain $A_C$}  &   \makecell[l]{Large-scale spatial \\ variation $A_C$} &  \makecell[l]{Small-scale IV \\ $A_{UC} = A - A_C$}   &  {Method}&   {$S_i$}&{Analysis Model} 
    \\
    \hline
    \hline
        \cite{dupont2022a} & \makecell[l]{thin-plate spline \\ basis} 
        
        & \makecell[l]{$\hat{g}(S)$ }
        & \makecell[l]{$A-\hat{g}(S)$}
         
        & 2SLS 
        & geos.
        & \makecell[l]{
        $Y_i = \beta_0 + \beta(A_i-\hat{g}(S_i))  + h(S_i) + \epsilon_i$, \\
        $h$ is also obtained \\ via a thin-plate spline with  \\ same df. as $g$} 
        \\
        \hline
        \cite{urdangarin2024simplified} 
        & \makecell[l]{$k+1$ eigenvectors \\ of spatial precision  \\matrix}
        & \makecell[l]{ $\sum_{i = n-k}^{n} v_iv_i^TA$}
        & $\sum_{i = 1}^{n-(k+1)} v_iv_i^TA$   & 2SLS &  areal
        & \makecell[l]{$Y_i \sim \text{Pois}(e_i R_i)$ \\ 
        $\log R = 1_n \alpha + (\sum_{i = 1}^{n-(k+1)} v_iv_i^TA)\beta + \theta$, \\ $\theta \sim \mathcal{N}(0_n, \Omega^{-1})$,\\
        $\Omega$ is a spatial precision matrix}
        \\
        \hline        
        \cite{keller2020a} (preadjustment of exposure) & \makecell[l]{TPS/Fourier/Wavelet\\ basis of dimension\\ $m$, $H_m$}
        &  \makecell[l]{ $ H_m(H_m^T H_m)^{-1} H_m^T A$ } 
        & \makecell[l]{$A-H_m(H_m^T H_m)^{-1} H_m^T A$  
        }  
        & 2SRI&   geos.
         &\makecell[l]{
        $Y = \beta_01_n$ \\ 
        \hspace{0.25cm} $+ \beta(A-H_m(H_m^T H_m)^{-1}H_m^T A) $ \\
        \hspace{0.5cm} $+ \gamma(H_m(H_m^T H_m)^{-1}H_m^T A ) + \epsilon$
        } 
        \\
        \hline        

        \cite{guan2022spectral} (sdiscrete-space) & \makecell[l]{$n-1$ eigenvectors \\ of the Graph\\  Laplacian }
        
        & \makecell[l]{$\sum_{i = 1}^{n-1} v_i v_i^T A$ } & \makecell[l]{$v_n v_n^T A$ }  & 2SRI &  areal
        &  \makecell[l]{$Y = \beta_0 1_n + \sum_{i = 1}^n v_i v_i^T A \bigg(\sum_{l = 1}^L b_l B_l(\omega_i)\bigg) $ \\
        \hspace{0.25cm}$+ V + \epsilon$, $ V \sim \text{ CAR}$ \\
        $B_l(\omega_k)$ are spline basis functions \\
        with coefficients $b_l$.\\
        $\hat{\beta} = \sum_{l = 1}^L B_l(\omega_n) \hat{b}_l$.}
        \\
        \hline
        
        \cite{thaden2018structural} &Indicators $z_{1}, \ldots, z_{d}$ for $d$ regions  & \makecell[l]{$\sum_{k = 1}^d z_{k}\gamma_{1k}$ } & \makecell[l]{$A - \sum_{k = 1}^d z_{k}\gamma_{1k}$ }
         & double pred. &  areal
         & \makecell[l]{ 
        $Y_i - \sum_{k = 1}^d z_{ki} \gamma_{2k} $ \\
        \hspace{0.25cm}$= \beta(A_i - \sum_{k = 1}^d z_{ki} \gamma_{1k}) + \epsilon_i$}\\
        
        \hline
        \cite{wiecha2024two} & universal kriging  & \makecell[l]{$\hat{g}(S)$ } 
        & \makecell[l]{$A - \hat{g}(S)$ }  & double pred. &  geos.
        & \makecell[l]{$Y_i - \hat{h}(S_i) = \beta (A_i - \hat{g}(S_i)) + \epsilon_i$\\
         $\hat{h}$ denotes the estimated spatial trend in \\ $Y$ obtained through universal kriging}\\
        \hline
        
    \end{tabular}
    \caption{\footnotesize{
    Our framework unifies six methods for addressing spatial confounding bias by demonstrating that they are instrumental variable (IV) approaches. Central to our framework are four key assumptions, derived from an implicit decomposition of exposure into confounded and unconfounded components (Assumptions 1--5). Each method is further distinguished by two primary characteristics. }The first property is the spatial decomposition, which partitions the exposure into large-scale variation correlated with the endogenous error $(A_C)$ and small-scale variation uncorrelated with the error $(A_{UC})$, which serves as the instrument. To ensure the identifiability of $\beta$, it is crucial that the exposure includes variation not fully spanned by the spatial basis defining $A_C$ (Assumption 5). The second characteristic concerns the specific IV method used to exploit this decomposition in estimating $\beta$. These methods include: (1) two-stage least squares (2SLS), where $A_{UC}$ is substituted for exposure in the outcome regression; (2) two-stage residual inclusion (2SRI), where $A_{UC}$ is substituted for exposure in the outcome regression and $A_C$ is included as an additional covariate; or (3) double prediction (double pred.), where the outcome, after being residualized to remove its confounded variation, is regressed on $A_{UC}$. The remaining columns of this table are defined as follows. (1) `Spatial basis or method used to obtain $A_C$" refers to the basis or estimation approach used to obtain $A_C$. (5) $S_i$ denotes whether the method was originally constructed for areal or geostatistical spatial data. (6) ``Analysis model" describes the model that is used to analyze the observed data. 
    }
    \label{tab:connections-expanded}
\end{table}
\end{landscape}

\section{Causal identification}

\subsection{Comparison with \cite{imbens2009identification}}
Our causal identification results build upon the IV theory of \cite{imbens2009identification}, although their results were not developed in the context of unmeasured spatial confounding.  
For clarity, Table \ref{tab:comparison} juxtaposes our assumptions and propositions with those of \cite{imbens2009identification} when their identification approach is directly applied to our setting.

Mapping our notation to the notation used by \cite{imbens2009identification}, $A = X = X_1$ and $Z_1$ does not exist. The unmeasured confounder is $U = \epsilon$ and the instrument is $A_{UC} = Z$. Lastly, $A_C = \eta$.
\begin{table}[!ht]
    \centering
    \footnotesize
    \def\arraystretch{2}
    \begin{tabular}{p{1in}|p{2.9in}|p{2.5in}}
    & Proposed Methodology & \cite{imbens2009identification}\\
    \hline
       \multirow{ 4}{*}{Assumptions} & A6: $Y_i = Y_i(a) $ if $A_i = a$&   \\
       & A7: $Y_i(a) = m(a, X_i, U_i)$ & A7': $Y_i = m(A_i, U_i)$ \\
        & A8: $A_i = A_{UC_i}+ A_{C_i}$ & \makecell[l]{A8': $A_i = h({A_{UC}}_i, {A_C}_i)$,\\ $h$ strictly monotonic in its second  \\argument with probability $1$} \\
        & A9: $A_{UC_i} \indep (U_i, A_{C_i}) |X_i$& 
 A9': $A_{UC_i} \indep (U_i, A_{C_i})$ \\
        & & \makecell[l]{A10': $A_C$ is a continuously distributed\\
        scalar with CDF that is strictly \\ increasing on its support } \\
        \hline
        \multirow{ 3}{*}{Propositions} & P1:
    $A_i \indep  U_i|(X_i, {A_C}_i)$ &  P1': $A_i \indep U_i | V_i = F_{A|A_{UC}}(A_i|{A_{UC}}_i)$\\
    & P2: $Y_i(a) \indep A_i | (X_i, {A_C}_i)$ $\forall a \in \text{supp}(A)$ & P2': $m(a,U) \indep A | V$ 
    \end{tabular}
    \caption{Comparison of our proposed methodology with \cite{imbens2009identification}. To align with existing spatial confounding literature and facilitate causal inference, we introduce three modifications: (1) the use of potential outcomes notation; (2) a known, additive decomposition of exposure; and (3) additional conditioning on measured confounders.}
    \label{tab:comparison}
\end{table}

Our methodology differs from \cite{imbens2009identification} in three ways. First, we explicitly introduce potential outcomes notation to enable causal inference (A6). Second, \cite{imbens2009identification} relax our Assumption $8$ by assuming that $A = h(A_C, A_{UC})$ for some unknown function $h$ that is strictly monotonic in its second argument with probability $1$. We instead substitute $h$ with a simple additive function to align with the existing spatial confounding methods in the literature. Additionally, this avoids the need to estimate and adjust for $V = F_{A|A_{UC}}(A|A_{UC})$, because we can directly adjust for $A_C$. Consequently, A10' is no longer required. 

The third distinction is the inclusion of measured confounders, $X$, which allows us to replace the independence assumption (A9') with conditional independence (A9) of the instrument. We consider the latter to be more plausible in many contexts. 
The conditioning on $X$ is extended throughout the propositions, so that conditional ignorability is achieved by conditioning on both $A_C$ and $X$, rather than $A_C$ alone.

\subsection{Proofs of propositions and corollaries}
Below we present the proofs of Propositions 1--2 and Corollaries 2.1--2.2 under Assumptions 6--9. 

\noindent \textbf{Proposition 1}
\begin{proof}
For any bounded function $g$, Assumptions 8--9 imply
\begin{align*}
    \E(g(A)|A_C, X, U) &= \int g(a) dF_{A|A_C, X, U}(a)\\
    &= \int g(a_{uc} + A_C) dF_{A|A_C, X, U}(a_{uc})\\
    &= \int g(a_{uc} + A_C) dF_{A|A_C, X}(a_{uc})\\
    &= \int g(a) dF_{A|A_C, X}(a)\\
    &= \E(g(A)|A_C, X).
\end{align*}
Therefore, for any bounded function $f$, we have \begin{align*}
    \E(g(A) f(U)|A_C, X) &= \E(f(U)\E(g(A)|A_C, X, U)|A_C, X)\\
    &= \E(f(U)\E(g(A)|A_C, X)|A_C, X)\\
    &= \E(f(U)|A_C, X)\E(g(A)|A_C, X).
\end{align*} 
\end{proof}

\noindent \textbf{Proposition 2}
\begin{proof}
    Combining Assumption 7 with Proposition 1, the result follows.
\end{proof}

\noindent \textbf{Corollary 2.1}
    \begin{proof}
By Assumption 6 (consistency) and Proposition 2 (conditional ignorability),
\begin{align*}
    \E(Y(a)) &= \E(\E(Y(a)|X, A_C))\\
    &= \E(\E(Y(a)|A, X, A_C))\\
    &= \E(\E(Y|A=a, X, A_C))\\
    &= \int \E(Y|A=a, X = x, A_C= a_c) dF_{A_C, X}(a_c, x)
\end{align*}for all $a$ where the following positivity assumption is satisfied:
$$\text{supp}(A_C, X) = \text{supp}(A_C, X|A = a').$$ 
\end{proof}

\noindent \textbf{Corollary 2.2}
\begin{proof}
    By Assumption 6 (consistency) and Proposition 2 (conditional ignorability), \begin{align*}
        \frac{\E(Y(\min(A,c)))}{\E(Y(A))} &= \frac{\E(\E(Y(\min(A,c)|X, A_C)))}{\E(Y)} \\
        &= \frac{\E(\E(Y(\min(A,c)|A = \min(A,c), X, A_C)))}{\E(Y)} \\
        &= \frac{\E(\E(Y|A = \min(A,c), X, A_C)))}{\E(Y)} \\
        &= \frac{\int \E(Y \mid A = \min(a,c),  A_C=a_c, X= x) dF_{A, A_C, X}(a, a_c, x)}{\E(Y)}.
    \end{align*}
    if $a>c, (a,a_c,x) \in \text{supp}(A, A_C, X) \Rightarrow (c,a_c,x) \in \text{supp}(A,A_C,X)$.
\end{proof}

\subsection{Identification and estimation of alternative estimands}
Assumptions 6--9 also identify many other causal estimands with a modified positivity assumption. 

The shift estimand
$$\tau_\delta:= \E(Y(A + \delta) - Y(A))$$
represents the expected change in population-level outcomes if all units' exposures increased by $\delta$ \citep{gilbert2021causal}.

\noindent \textbf{Corollary 2.3} (Identification of the shift estimand) Under Assumptions 6--9, $\tau_\delta$ is identified as
\begin{align*}
    \tau_\delta 
    &=  \E(\E(Y|A + \delta, X, A_C))-\E(Y)\\
    &= \int \E(Y|A = a + \delta, X = x, A_C = a_c) dF_{A, A_C, X}(a,a_c,x) - \E(Y)
\end{align*}
for values of $\delta$ where the following positivity assumption is satisfied: 
$$(a, a_C, x)\in \text{supp}(A, A_C, X) \implies (a+\delta, a_C, x) \in \text{supp}(A, A_C, X).$$
\begin{proof}
    By Assumption 6 (consistency), Proposition 2 (conditional ignorability), and the positivity assumption, \begin{align*}
    \tau_\delta = \E(Y(A + \delta)-Y(A)) &= \E(\E(Y(A + \delta)|X, A_C))-\E(Y) \\
    &=  \E(\E(Y|A + \delta, X, A_C))-\E(Y)\\
    &= \int \E(Y|A = a + \delta, X = x, A_C = a_c) dF_{A, A_C, X}(a,a_c,x) - \E(Y).
\end{align*}
\end{proof}

More generally, Assumptions 6--9 identify the effects of modified treatment policies, where treatment is assigned as a function 
$q(A,X)$ of the observed exposure and covariates \citep{haneuse2013estimation}. The shift estimand is a special case with $q(A,X)=A+\delta$. The exposure-response curve is a special case with $q(A, X) = a$. The truncated exposure effect is a special case with $q(A,X)=\min(A, c)$.

\noindent \textbf{Corollary 2.4} (Identification of the effects of modified treatment policies) Suppose that Assumptions 6--9 hold. Further suppose that the following positivity assumption holds: $$(a,a_c, x) \in \text{supp}(A, A_C, X) \implies (q(a, x), a_c, x) \in \text{supp}(A, A_C, X).$$ Then $\E(q(A,X))$ is identified as 
\begin{align*}
    \E(Y(q(A, X))) &= \E(\E(Y|q(A,X), X, A_C)).
\end{align*}
\begin{proof} 
Combining Assumption 6 (consistency), Proposition 2 (conditional ignorability), and the positivity assumption, the result follows. 
\end{proof}

In summary, Assumptions 6--9 of our instrumental variables framework yield conditional ignorability, conditioning on measured covariates and the smooth spatial trend in exposure $A_C$. This allows for identification of many causal effects. Once causal identification has been established, estimation can proceed in several ways. Briefly, we mention three approaches that can be applied to estimate the effects of modified treatment policies. The outcome regression estimator is consistent if the conditional mean outcome model  $\E(Y|A, X, A_C)$ is correctly specified. The inverse probability weighting estimator is consistent under correct specification of the treatment density $\pi(A|X, A_C)$. The doubly robust estimator remains consistent if either the outcome model or the treatment model is correctly specified, offering additional protection against misspecification. For further details, see \citep{haneuse2013estimation}.

\section{   Doubly robust estimation of the truncated exposure effect}
In the following paragraphs we provide further details on our estimation procedure of the truncated exposure effect. We apply this same procedure across all approaches (baseline, oracle, IV-TPS, IV-GL, spatialcoord, IV-TPS+spatialcoord, and IV-GL +spatialcoord); the only distinction between them lies in the confounding adjustment sets, i.e., the variables included in $X$.

First, we describe estimation of $$\nu(c):= \E(\E(Y|A = c, X, A_C)|A \geq c) =  \E_\mathcal{P}(\E(Y|A = c, X, A_C))$$where $\mathcal{P}$ is the population with $A \geq c$. To protect against model misspecification, we use a doubly robust mapping $\xi((X, A_C, A, Y); \pi, \mu)$ whose conditional expectation under exposure equals $\nu(c)$, if either the conditional exposure density $\pi(a|x, a_c) = f_{A|X, A_C}(a| X = x, A_C = a_c)$ or outcome mean model $\mu(x, a_c, a) = \E(Y|X = x, A_C = a_c, A = a)$ are correctly specified in the population $\mathcal{P}$, following \citep{kennedy2017non}. In particular, defining $$\xi((X, A_{C}, A, Y); \pi, \mu) = \frac{Y - \mu(X, A_{C}, A)}{\pi(A|X, A_{C})}\int_\mathcal{P} \pi(A| x, a_c) dF_{ X, A_C}(x, a_c) + \int_\mathcal{P} \mu(x, a_c, A) dF_{ X, A_C}(x, a_c)$$we have $$\E_\mathcal{P}(\xi((X, A_{C}, A, Y); \overline{\pi}, \overline{\mu})|A = c)  = \E_\mathcal{P}(\E(Y|A = c, X, A_C)) = \nu(c)$$ if either $\overline{\pi} = \pi$ or $\overline{\mu} = \mu$. This suggests estimating $\pi$ and $\mu$ using off-the-shelf non-parametric regression or machine learning methods and then regressing the pseudo-outcome $$\hat{\xi}((X, A_C, A, Y); \hat{\pi}, \hat{\mu}) = \frac{Y - \hat{\mu}(X, A_C, A)}{\hat{\pi}(A|X, A_C)} \frac{1}{n}\sum_{i = 1}^n \hat{\pi}(A|X, A_{C}) + \frac{1}{n}\sum_{i = 1}^n \hat{\mu}(X, A_{C}, A)$$ on exposure $A$, restricting all estimation to units $i$ with $A_i \geq c$. We conducted all estimation with the R package \verb|npcausal|. For estimation of $\pi$ and $\mu$, we use a combination of candidate learners within Superlearner, including generalized additive models (SL.gam), generalized linear models (SL.glm), mean regression (SL.mean), and an interaction model (SL.glm.interaction). For the pseudo-outcome regression, we employ a local linear kernel estimator using bandwidth selection. 

Ultimately, $\hat{\nu}(c)$ is the estimated effect curve $\hat{\nu}(a)$ evaluated at $a = c$. 
Minor modifications were made to the \verb|ctseff| code in the \verb|npcausal| package to improve functionality. These changes can be found here: \url{https://github.com/ehkennedy/npcausal/issues/6}. 

In our data application, we encountered near-violations of the positivity assumption. In particular, a small number of observations yielded values of the estimated conditional exposure density $\hat{\pi}(A|X, A_C)$ that were nearly zero, leading to extreme values of the pseudo-outcome $\hat{\xi}$. To address this issue, we constrained $\hat{\xi}$ to lie within the range of the observed outcomes, as recommended in the Supplementary Material of \cite{kennedy2017non}.

We now return to estimation of the truncated exposure effect. The identifying functional in Proposition 3 can be rewritten as \begin{align*}
    \psi := \frac{\E(Y(\min(A,c)))}{\E(Y(A))} &= \frac{\E(\E(Y|A = \min(A,c), X, A_C)))}{\E(Y)}\\
    &= \frac{\E(\E(Y|A = c, X, A_C))|A \geq c)\mathbb P(A \geq c) + \E(Y|A < c) \mathbb P(A < c)}{\E(Y)}.
\end{align*}
Our proposed estimator of the truncated exposure effect thus takes the form \begin{align*}
    \hat{\psi} &= \frac{\hat{\nu}(c) \bigg(\frac{1}{n}\sum_{i= 1}^n I(A_i \geq c)\bigg) + \frac{\sum_{A_i < c} Y_i}{\sum_{i= 1}^n I(A_i < c)}\bigg(\frac{1}{n}\sum_{i= 1}^n I(A_i < c)\bigg)}{\frac{1}{n}\sum_{i= 1}^n Y_i},
\end{align*}where $\hat{\nu}(c)$ is the estimator of $\nu(c) =  \E(\E(Y|A = c, X, A_C)|A \geq c)$ described above. If either $\bar{\pi} = \pi$ or $\bar{\mu} = \mu$, $\hat{\psi} \stackrel{p}{\rightarrow} \psi$. We therefore refer to $\hat{\psi}$ as ``doubly robust".

\subsection{  Uncertainty quantification}
To assess the variability of our estimates in our data application, we apply the Delta Method. Recall that
\begin{align*}
\psi &= \frac{\E(\E(Y|A = c, X, A_C))|A \geq c)\mathbb P(A \geq c) + \E(Y|A < c) \mathbb P(A < c)}{\E(Y)},\\
    \hat{\psi} &= \frac{\hat{\nu}(c) \bigg(\frac{1}{n}\sum_{i= 1}^n I(A_i \geq c)\bigg) + \frac{\sum_{A_i < c} Y_i}{\sum_{i= 1}^n I(A_i < c)}\bigg(\frac{1}{n}\sum_{i= 1}^n I(A_i < c)\bigg)}{\frac{1}{n}\sum_{i= 1}^n Y_i},
\end{align*}where $\hat{\nu}(c)$ is the estimator of $\nu(c) =  \E(\E(Y|A = c, X, A_C)|A \geq c)$ previously described.

Let $\theta_1 = \E(\E(Y|A = c, X, A_C))|A \geq c)$, $\theta_2 = \mathbb P(A < c)$, $\theta_3 = \E(Y|A < c)$, and $\theta_4 = \E(Y)$. 

The first parameter $\theta_1$ has an estimated efficient influence function $\hat{\varphi}_1(A_i, X_i, Y_i)$ given by \cite{kennedy2017non} for units with $A_i \geq c$, and $0$ otherwise. We refer the reader to Section 3.4 of \cite{kennedy2017non} for its specific form. The remaining three parameters have estimated influence functions \begin{align*}
    \hat{\varphi}_2(A_i, X_i, Y_i) &= I(A_i < c) - \frac{1}{n}\sum_{i= 1}^n I(A_i < c)\\
    \hat{\varphi}_3(A_i, X_i, Y_i) &= \begin{cases} Y_i - \frac{\sum_{A_i < c} Y_i}{\sum_{i= 1}^n I(A_i < c)} & A_i < c\\
    0 & A_i \geq c
        \end{cases}\\
    \hat{\varphi}_4(A_i, X_i, Y_i) &= Y_i - \bar{Y}
\end{align*}respectively.

An estimate of the $4 \times 4$ covariance matrix of the influence functions is simply given by $\hat{\Sigma} = \widehat{\text{Cov}}(\hat{\varphi}_1, \hat{\varphi}_2, \hat{\varphi}_3, \hat{\varphi}_4)$, where $\widehat{\text{Cov}}$ denotes empirical covariance. 

Since $\psi = \frac{\theta_1(1-\theta_2) + \theta_3 \theta_2}{\theta_4}$ has a gradient of $$\nabla = \bigg(\frac{1-\theta_2}{\theta_4}, \frac{-\theta_1 + \theta_3}{\theta_4}, \frac{\theta_2}{\theta_4}, -\frac{\theta_1(1-\theta_2) + \theta_2\theta_3}{\theta_4^2}\bigg)^T,$$ we have 
$$\sqrt{n}(\hat{\psi} - \psi) \stackrel{d}{\rightarrow} \mathcal{N}(0, \nabla^T \Sigma \nabla)$$
by the Delta Method and a 95\% CI for $\hat{\psi}$ is given by
$\hat{\psi} \pm 1.96 \sqrt{\frac{\hat{\nabla}^T \hat{\Sigma} \hat{\nabla}}{n}}$.

\section{  Additional simulation details}
\label{suppsec:sim}
\subsection{  Spatial confounding approaches}
Table \ref{tab:method-summary} summarizes the seven spatial confounding approaches compared in our simulation study and data application. We additionally introduce two methods, titled ``trueIV'' and ``trueIV+spatialcoord'', that adjust for the true confounded exposure component $A_C$ instead of estimating it with a thin plate spline or Graph Laplacian basis. As described in the Supplementary Material, we apply the same estimation procedure across all approaches, and the only distinction between them lies in the confounding adjustment sets, i.e. the variables that are adjusted for. 

\begin{table}[!ht]
\centering
\renewcommand{\arraystretch}{1.15}
\begin{tabular}{c|c}
Method & Confounding adjustment set \\
\hline
oracle  & $U$ \\
baseline  & -- \\
spatialcoord & latitude, longitude \\
trueIV & $A_C$ \\
IV-TPS & $\widehat{A}^{\text{TPS}}_C = P_{\text{TPS}} A$ \\
IV-GL & $\widehat{A}^{\text{GL}}_C = P_{\text{GL}} A$ \\
trueIV+spatialcoord & $A_C$, latitude, longitude \\
IV-TPS+spatialcoord & $\widehat{A}^{\text{TPS}}_C$, latitude, longitude \\
IV-GL+spatialcoord & $\widehat{A}^{\text{GL}}_C$, latitude, longitude
\end{tabular}
\caption{
Overview of spatial confounding approaches assessed in simulations. 
Here $P_{\text{TPS}} = B^{\text{TPS}} (B^{\text{TPS}}{}^\top B^{\text{TPS}})^{-1} B^{\text{TPS}}{}^\top$ 
and $P_{\text{GL}} = B^{\text{GL}} (B^{\text{GL}}{}^\top B^{\text{GL}})^{-1} B^{\text{GL}}{}^\top$ 
are the projection matrices onto the thin plate spline and Graph Laplacian bases, respectively. 
$B^{\text{TPS}} \in \mathbb{R}^{n \times K}$ is the thin plate spline design matrix built from locations 
$\{(\text{lat}_i, \text{long}_i)\}$ and $B^{\text{GL}} \in \mathbb{R}^{n \times K}$ contains the $K$ eigenvectors 
associated with the smallest nonzero eigenvalues of the Graph Laplacian, with $K = 35$.
}
\label{tab:method-summary}
\end{table}

Our implementation of the spatial coordinates (spatialcoord) method corresponds most closely to the ``DML Spatial location'' variant proposed by \cite{gilbert2021causal}, modified as described in the previous section to target the truncated exposure effect.

\subsection{  Confounding mechanisms}
\begin{enumerate}
    \item 
    The first confounding mechanism generates: 
\begin{align*}
    \begin{pmatrix} A_{UC} \\
    A_C\\
    U
    \end{pmatrix} &\sim \mathcal{N}\bigg\{\begin{pmatrix} (0.1)1_n \\ (-0.2)1_n \\ (0.3)1_n \end{pmatrix}, \begin{pmatrix} R(\theta_{A_{UC}}) & 0 & 0 \\
    0 & R(\theta_{A_{C}}) & 0.95 R(\theta_{A_{C}})\\
    0 & 0.95 R(\theta_{A_{C}}) & R(\theta_{A_{C}})
    \end{pmatrix}\bigg\},
\end{align*} 
\noindent with $\theta_{A_{UC}} = 0.01$ and $\theta_{A_{C}} = 0.5$, so that the spatial range or scale of the unconfounded component of exposure is much smaller than that of the confounded component. 
    \item  
    The second confounding mechanism generates spatially correlated random fields using the bivariate Leroux conditional autoregressive (CAR) model. Let $W$ denote the $n \times n$ adjacency matrix with $W_{ij} = 1$ if county $i$ borders county $j$ and $0$ otherwise. Let $D$ be the diagonal degree matrix whose $i$th diagonal entry is equal to $D_{ii} = \sum_{j = 1}^n W_{ij}$. Define $Q =(1-\rho)I_n + \rho (D - W)$ with $\rho = 0.8$. Define the $2\times 2$ cross covariance matrix $$\Sigma = \begin{pmatrix} 1 & 0.7 \\ 0.7 & 1\end{pmatrix}$$ with inverse $R$. A bivariate spatial process for $(U, A_C)^T$ is then generated as $$\begin{pmatrix} U \\ A_C \end{pmatrix} \sim \mathcal{N}\bigg(0_{2n}, (R \otimes Q)^{-1}\bigg),$$ and $A_{UC} \sim \mathcal{N}(0_n, I_n).$
    \item  
    The third confounding mechanism uses the same Gaussian process as the first, applied independently across states.
    \item  
    The fourth confounding mechanism is constructed such that two unmeasured confounders jointly induce bias. In particular, data is generated as:
\scriptsize
\begin{align*}
    \begin{pmatrix} A_{UC} \\
    A_C\\
    U_1 \\ 
    U_2
    \end{pmatrix} &\sim \mathcal{N}\bigg\{\begin{pmatrix} (0.1)1_n \\ (-0.2)1_n \\ (0.3)1_n \\ (-0.1)1_n \end{pmatrix}, \begin{pmatrix} R(\theta_{A_{UC}}) & 0 & 0 & 0\\
    0 & R(\theta_{A_{C}}) & \rho_1  R(\theta_{A_{C}}) & \rho_2 R(\theta_{A_{C}})\\
    0 & \rho_1 R(\theta_{A_{C}}) & \rho_1^2  R(\theta_{A_{C}}) + (1-\rho_1^2) R(\theta_{U_1}) & \rho_1 \rho_2 R(\theta_{A_{C}})\\
    0& \rho_2 R(\theta_{A_{C}})& \rho_1 \rho_2 R(\theta_{A_{C}}) & \rho_2^2  R(\theta_{A_{C}}) + (1-\rho_2^2) R(\theta_{U_2})
    \end{pmatrix}\bigg\},
\end{align*} 
\normalsize
\noindent with $\theta_{A_{UC}} = 0.01$, $\theta_{A_{C}} = 0.5$, $\theta_{U_1} = 0.5$, and $\theta_{U_2} = 0.3$ and $R$ is a Mat\'ern spatial correlation function with smoothness parameter $\nu = 2$. The small range for $A_{UC}$ implies that the unconfounded component of exposure varies at a much finer spatial scale than the confounded component. The correlation parameters $\rho_1,\rho_2$ determine how strongly $U_1$ and $U_2$ correlate with the confounded exposure component $A_C$.
The additive terms $(1-\rho_j^2)R(\theta_{U_j})$ in the diagonal blocks ensure that each confounder also exhibits its own spatial pattern, with range parameters $\theta_{U_1},\theta_{U_2}$ governing the scale of this idiosyncratic variation. Exposure is subsequently generated as $A = A_C + A_{UC}$.

The linear and non-linear outcome models are formulated differently in order to incorporate two unmeasured confounders. Specifically, the linear outcome model is $Y_i \sim \mathcal{N}(-0.5 - U_{1i} + A_i - 0.5 A_i U_{1i} - 0.75 A_{i} U_{2i}, 1)$. The non-linear outcome model is $Y_i \sim \mathcal{N}(-0.5 - 0.5U_{1i} + \tanh (1.5A_i) - 0.2 U_{2i}\tanh A_i +0.1 (\tanh A_i)^2,1)$.

    \item  
    The fifth confounding mechanism is the same as the first, but the relative spatial scales of $A_{UC}$ and $A_C$ are reversed by setting $\theta_{A_{UC}} = 0.1$ and $\theta_{A_{C}} = 0.01$. 
    \item
    The sixth confounding mechanism is the same as the first, except that the unconfounded variation is smoother in space by setting $\theta_{A_{UC}} = 0.05$. In this case, the IV methods adjust for the residuals instead of the predicted values of the thin plate spline and Graph Laplacian regressions.
    \item 
    The seventh confounding mechanism sets $U_i = \sin(2 \pi \text{lat}_i\text{long}_i) +  \text{lat}_i +  \text{long}_i$, for normalized spatial coordinates $(\text{lat}_i,\text{long}_i)$, following a simulation analysis by \cite{gilbert2021causal}. The two exposure components are generated as $A_{C_i} \sim \mathcal{N}(U_i, 0.1^2)$ and $A_{UC_i} \sim \mathcal{N}(0, 1)$. 
\end{enumerate}

\subsection{  True truncated exposure effects}
For each combination of confounding mechanism and outcome model, we approximate the true truncated exposure effect $\tau^*$  with the mean of the oracle estimates.
\renewcommand{\arraystretch}{0.6}
\begin{center}\begin{tabular}{ccc}
  \hline
Confounding Mechanism & Outcome Model & $\tau^*$ \\ 
  \hline
1 & linear & 1.1012 \\ 
  1 & nonlinear & 1.0592 \\ 
  2 & linear & 1.2597 \\ 
  2 & nonlinear & 1.0944 \\ 
  3 & linear & 1.1036 \\ 
  3 & nonlinear & 1.0742 \\ 
  4 & linear & 1.0075 \\ 
  4 & nonlinear & 1.0827 \\ 
  5 & linear & 1.0822 \\ 
  5 & nonlinear & 1.0377 \\ 
  6 & linear & 1.1019 \\ 
  6 & nonlinear & 1.0611 \\ 
  7 & linear & 1.0763 \\ 
  7 & nonlinear & 1.1959 \\ 
   \hline
\end{tabular}\end{center}

\begin{figure}[!ht]
    \centering
    \includegraphics[width=\linewidth]{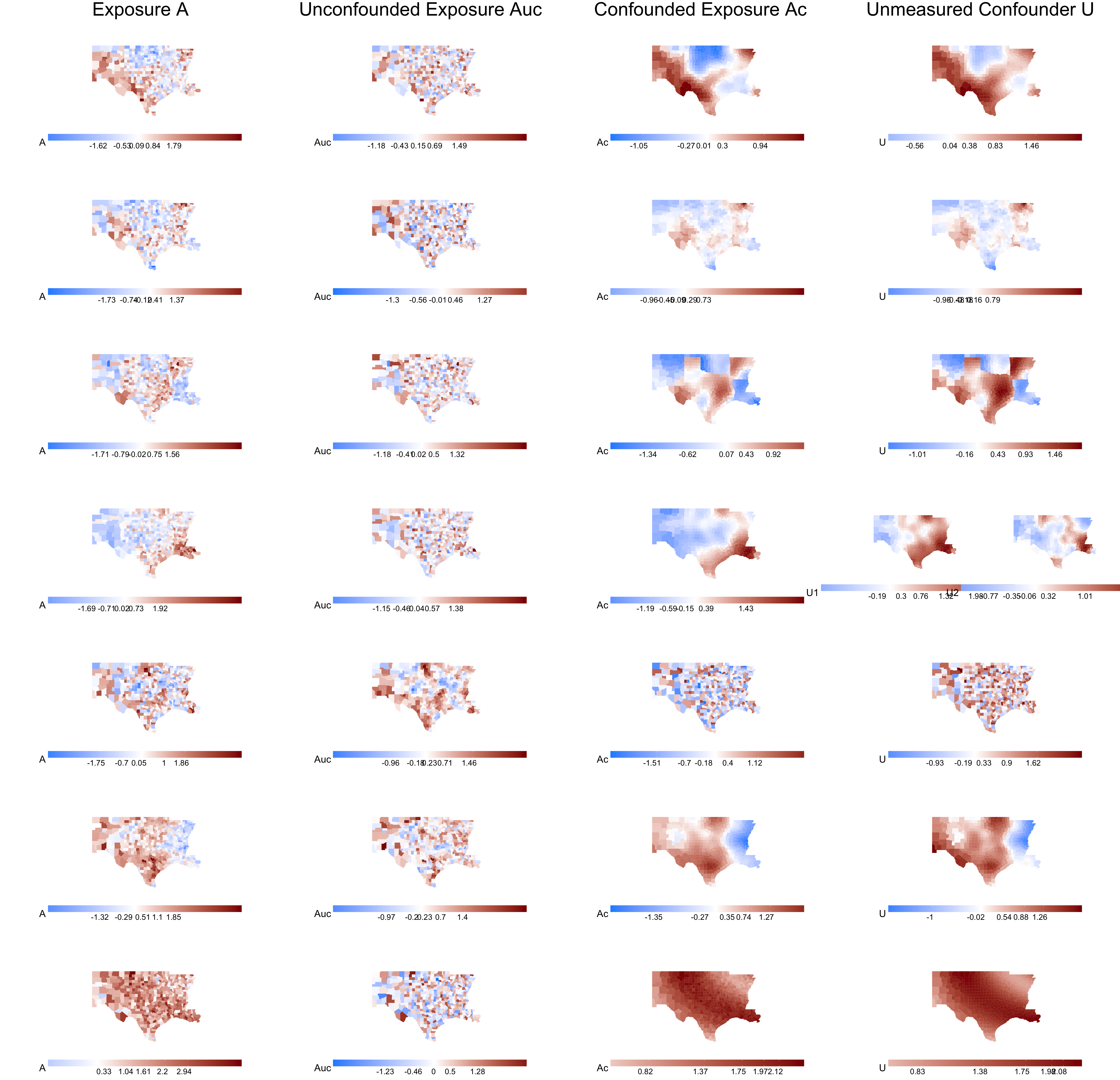}
    \caption{One observation of $(A, A_{UC}, A_C, U)$ for each of the seven confounding mechanisms. 
    }
    \label{fig:conf-mech}
\end{figure}

Figure \ref{fig:conf-mech} plots one observation of $(A, A_{UC}, A_C)$ for each of the three confounding mechanisms. 
\vspace{0.5in}

\subsection{  Expanded simulation results}
\begin{landscape}
    
\begin{figure}[!ht]
    \centering
    \includegraphics[width=\linewidth]{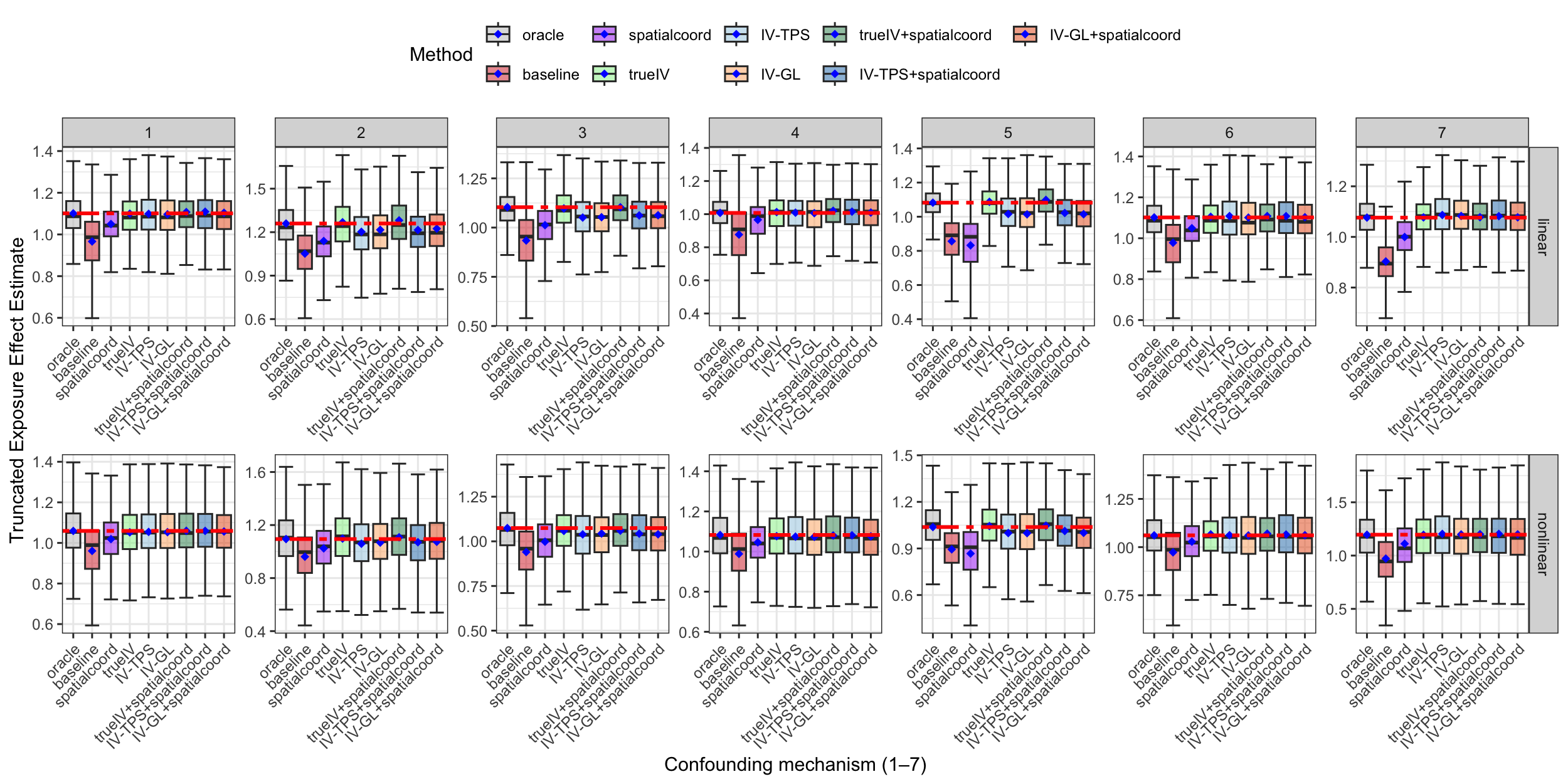}
    \caption{Estimates of the truncated exposure effect across $1000$ simulations for each combination of outcome model (linear or non-linear) and confounding mechanism (1, 2, 3, 4, 5, 6, or 7) for each of the nine methods. The blue diamonds indicate the means of the estimates, and the dashed red lines correspond to the means of the oracle estimates.}
    \label{fig:boxplot}
\end{figure}
\end{landscape}

Table \ref{tab:combined_tall_results} presents the absolute bias and RMSE of the truncated exposure effects estimate across the $1000$ simulations for each of the fourteen confounding scenarios and nine methods relative to the oracle mean. Figure \ref{fig:boxplot} provides boxplots of the estimates.
\begin{landscape}
\centering
\begin{table}
\small
\begin{tabular}{p{2.5cm}p{2.5cm}p{1.5cm}p{1.5cm}p{1.5cm}p{1.5cm}p{1.5cm}p{1.5cm}p{1.5cm}p{1.5cm}p{1.5cm}}
\makecell[l]{Confounding \\Mechanism} & \makecell[l]{Outcome \\Model} & Oracle & Baseline & \makecell[l]{Spatial\\coord} & trueIV  & IV-TPS & IV-GL & \makecell[l]{trueIV\\+spatial\\coord} & \makecell[l]{IV-TPS\\+spatial\\coord} & \makecell[l]{IV-GL\\+spatial\\coord} \\
\hline
\multicolumn{11}{c}{Absolute bias} \\
\hline
1 & linear & 0.00 & 13.49 & 5.11 & \textbf{0.15} & 0.22 & 0.47 & 0.62 & 0.79 & 0.29 \\ 
  1 & nonlinear & 0.00 & 9.86 & 4.05 & 0.69 & 0.48 & 0.61 & 0.06 & \textbf{0.01} & 0.29 \\ 
  2 & linear & 0.00 & 20.77 & 12.08 & \textbf{0.84} & 5.73 & 4.35 & 2.17 & 4.49 & 3.55 \\ 
  2 & nonlinear & 0.00 & 13.31 & 7.03 & \textbf{0.36} & 3.54 & 2.68 & 1.35 & 2.38 & 2.06 \\ 
  3 & linear & 0.00 & 16.93 & 9.13 & 0.63 & 5.27 & 5.09 & \textbf{0.05} & 4.20 & 4.10 \\ 
  3 & nonlinear & 0.00 & 13.41 & 7.64 & 1.59 & 3.53 & 3.04 & \textbf{1.44} & 3.14 & 3.49 \\ 
  4 & linear & 0.00 & 13.09 & 4.19 & 0.50 & 0.15 & \textbf{0.15} & 1.31 & 0.76 & 0.21 \\ 
  4 & nonlinear & 0.00 & 9.43 & 3.69 & 0.62 & 0.83 & 0.94 & 0.26 & \textbf{0.16} & 0.71 \\ 
  5 & linear & 0.00 & 22.58 & 24.98 & \textbf{0.50} & 6.64 & 6.85 & 1.63 & 6.03 & 6.78 \\ 
  5 & nonlinear & 0.00 & 14.35 & 17.07 & \textbf{0.63} & 3.86 & 3.92 & 0.90 & 2.53 & 3.73 \\ 
  6 & linear & 0.00 & 12.38 & 5.28 & 0.31 & 0.67 & 0.09 & 0.65 & 0.62 & \textbf{0.08} \\ 
  6 & nonlinear & 0.00 & 8.82 & 3.30 & 0.50 & 0.07 & \textbf{0.05} & 0.88 & 0.43 & 0.08 \\ 
  7 & linear & 0.00 & 17.32 & 7.66 & \textbf{0.09} & 1.03 & 0.67 & 0.17 & 0.57 & 0.10 \\ 
  7 & nonlinear & 0.00 & 22.40 & 8.47 & \textbf{0.14} & 0.74 & 0.26 & 0.45 & 0.77 & 0.33 \\ 
  \hline
\multicolumn{11}{c}{RMSE} \\
\hline
1 & linear & 11.98 & 21.44 & 13.33 & 13.05 & 14.32 & 14.03 & 12.71 & 13.45 & \textbf{12.49} \\ 
  1 & nonlinear & 17.95 & 20.37 & \textbf{16.18} & 19.67 & 19.95 & 20.31 & 18.44 & 19.45 & 17.55 \\ 
  2 & linear & 18.80 & 31.40 & 22.12 & 21.68 & 24.97 & 24.83 & \textbf{20.78} & 23.15 & 21.76 \\ 
  2 & nonlinear & 27.91 & 29.18 & \textbf{26.93} & 30.77 & 29.37 & 31.73 & 28.64 & 28.61 & 27.26 \\ 
  3 & linear & 17.50 & 23.84 & 16.32 & 12.39 & 17.44 & 18.72 & 11.60 & 14.09 & \textbf{13.88} \\ 
  3 & nonlinear & 50.49 & 22.79 & 18.94 & \textbf{18.04} & 53.19 & 59.61 & 17.88 & 19.46 & 18.81 \\ 
  4 & linear & 14.12 & 24.90 & 17.19 & 17.03 & 17.39 & 17.74 & 15.90 & \textbf{15.48} & 16.73 \\ 
  4 & nonlinear & 22.39 & 20.83 & \textbf{17.88} & 21.71 & 22.61 & 21.14 & 20.45 & 20.88 & 20.98 \\ 
  5 & linear & 11.14 & 27.73 & 31.25 & 13.86 & 17.25 & 17.59 & \textbf{12.79} & 14.65 & 16.24 \\ 
  5 & nonlinear & 20.95 & 22.04 & 26.95 & 22.07 & 19.90 & 21.11 & 20.51 & \textbf{17.62} & 19.19 \\ 
  6 & linear & 11.12 & 20.73 & 13.32 & 12.35 & 16.02 & 16.35 & \textbf{12.13} & 14.41 & 13.48 \\ 
  6 & nonlinear & 18.27 & 20.60 & 19.55 & \textbf{18.04} & 20.32 & 21.62 & 18.24 & 21.09 & 21.16 \\ 
  7 & linear & 9.87 & 20.16 & 15.28 & \textbf{10.00} & 11.95 & 12.60 & 10.97 & 12.91 & 13.13 \\ 
  7 & nonlinear & 28.66 & 34.27 & 36.34 & \textbf{28.42} & 32.50 & 34.69 & 31.13 & 36.03 & 36.49 \\ 
\end{tabular}
    \caption{Absolute bias and root mean squared error for each of the fourteen data-generating scenarios relative to the oracle mean. RMSE, root mean squared error; spatialcoord, spatial coordinates. All values have been multiplied by $10^2.$ }
\label{tab:combined_tall_results}
\end{table}
\end{landscape}

\section{   Data application details}
\label{suppsec:dataapp}
\subsection{  Dataset characteristics}
Table \ref{tab:dataset} provides a description of the zip code-level dataset $(n = 33,255)$ and data sources. To ensure numerical stability, the analysis was restricted to zip codes with more than 10 person-years contributing to the Medicare cohort. As a result, 669 zip codes (1.9\% of the original 33,464) were excluded.

\begin{table}
\small
\renewcommand{\arraystretch}{1.38}
    \begin{tabular}{p{2cm}| p{4.5cm} p{3cm} p{6cm}}
    & Variables & Mean(sd) & Data Source\\
    \hline
    Exposure & Long-term average to PM$_{2.5}$ during 2001--2010 ($\mu g/m^3)$ &  10.582 (2.947) & Daily estimates of PM$_{2.5}$ at the $1 \text{km} \times 1 \text{km}$ grid level obtained from a machine learning model combining ground, satellite and reanalysis data and subsequently aggregated to zip code-level using area-weighting \citep{di2017air}\\
    \hline
    Outcome & All-cause mortality rate among Medicare enrollees age $\geq 65$ during 2011--2016 $(1/\text{years})$ &  0.045 (0.014) & Medicare claims data, obtained from the Centers for Medicare and Medicaid Services \\
    \hline
    \multirow{14}{1.5cm}{Measured confounders} & proportion of Hispanic residents & 0.073 (0.144) & \multirow{8}{5cm}{U.S. Decennial Census, American Community Survey} \\
    & proportion of Black residents & 0.085 (0.167) & \\
    & median household income (\$) & 41104.241 (17083.639) & \\
    & median home value (\$) & 112029.075 (90828.365) & \\ 
    & proportion of residents in poverty & 0.110  (0.103) & \\
    & proportion of residents with high school diploma as highest level of education & 0.378 (0.188) & \\
    & population density (people/mi$^2$) & 1431.382 (4655.331) & \\
    & proportion of residents that own their house & 0.732 (0.168) & \\ \cline{2-4}
    & average body mass index & 26.926 (1.104) & \multirow{2}{6.5cm}{Centers for Disease Control and Prevention’s Behavioral Risk Factor Surveillance System}\\
    & proportion of smokers & 0.482 (0.075) & \\ \cline{2-4}
    & average maximum daily temperature in summer (K) & 301.938 (3.903) & \multirow{4}{5cm}{gridMET via Google Earth Engine}\\
    & average maximum daily temperature in winter (K) & 282.118 (6.823) &\\
    & average relative humidity in summer (\%) & 91.105 (10.239) &\\
    & average relative humidity in winter (\%) & 86.907 (8.011) & \\ 
    \end{tabular}
    \caption{Description of zip code-level dataset ($n = 33,255$) and data sources.}
    \label{tab:dataset}
\end{table}

\subsection{  Evaluation of truncated exposure effect estimates}
We compare the truncated exposure estimates from each method to the oracle estimate by computing the average Hausdorff distance between their associated confidence intervals across different cutoff values. For a given cutoff value, let $I_1 = (a_1,b_1)$ and $I_2 = (a_2,b_2)$ denote the confidence intervals from oracle and the method under consideration, respectively. The Hausdorff distance between $I_1$ and $I_2$ is defined as \begin{align*}
    d_H(I_1, I_2) = \max \{\sup_{x \in I_1} d(x,I_2), \sup_{y \in I_2} d(I_1, y)\} = \max \{|a_1 - a_2|, |b_1 - b_2|\}.
\end{align*}Hausdorff distance is averaged across cutoff values to assess overall alignment with the oracle estimates.

\begin{center}
\small
\singlespacing
\renewcommand{\arraystretch}{1.3}
    \begin{tabular}{l|p{1.5cm}p{1.5cm}p{1.75cm}p{1.75cm}p{1.75cm}p{1.75cm}}
  Cutoff  & baseline& \makecell[l]{spatial\\coord} & \makecell[l]{IV-TPS}& \makecell[l]{IV-GL} & \makecell[l]{IV-TPS\\+spatial\\coord} & \makecell[l]{IV-GL\\+spatial\\coord} \\
  \hline
  $6\mu g /m^3$ & 1.84 & 1.78 & 2.37 & 2.40 & \textbf{0.85} & 1.35 \\ 
  $7\mu g /m^3$ & 1.66 & 0.75 & 0.78 & 0.77 & 0.55 & \textbf{0.36} \\ 
  $8\mu g /m^3$ & 2.28 & 0.72 & 1.03 & 1.15 & 1.07 & \textbf{0.19} \\ 
 $9\mu g /m^3$ &  0.56 & 0.18 & \textbf{0.18} & 0.20 & 0.29 & 0.32 \\ 
  $10\mu g /m^3$ &  0.20 & 0.06 & 0.17 & 0.16 & 0.31 & \textbf{0.05} \\ 
 $11\mu g /m^3$&  \textbf{0.02} & 0.10 & 0.44 & 0.13 & 0.32 & 0.43 \\ 
  $12\mu g /m^3$ &  0.26 & 0.21 & 0.20 & 0.19 & 0.13 & \textbf{0.01} \\ 
  \hline
  Average &  0.97 & 0.54 & 0.74 & 0.71 & 0.50 & \textbf{0.39} \\ 
    \end{tabular}
\end{center}

\subsection{  Estimating the exposure-response curve between air pollution and mortality}
Below we estimate the exposure-response curve between long-term average air pollution and all-cause mortality in the United States using the zip code-level dataset ($n = 33,255$) described above. We employ seven different confounding adjustments within the doubly robust estimation method by \cite{kennedy2017non}, as in the analysis of the truncated exposure effect. For each estimation strategy, we calculate $\E(Y(a))$ at $100$ equally spaced values of $a$ within the percentile range $(2.5\%, 97.5\%)$ of exposure.

The first panel of Figure \ref{fig:ercs} shows the seven estimated curves. Comparing the second panel with the third, we observe that the baseline curve deviates significantly from the oracle curve for exposure values $6-9 \mu g/m^3$, as the oracle curve falls outside the confidence interval of the baseline and vice versa. Moreover, the oracle confidence intervals are somewhat narrower than those produced by the baseline. These findings indicate that unmeasured spatial confounding is indeed present by omitting temperature and humidity variables. 

The spatial coordinates approach, IV-TPS, IV-GL, IV-TPS+spatialcoord, and IV-GL+ spatialcoord all reasonably approximate the oracle curve estimate and appropriately quantify its uncertainty. Spatial coordinates, IV-TPS, and IV-GL slightly underestimate the oracle curve for exposure values $\leq 9 \mu g/ m^3$. IV-TPS+spatialcoord and IV-GL+spatialcoord produce estimates that most closely align with the oracle curve.

All estimated exposure-response curves suggest a statistically significant harmful effect of long-term average exposure to PM$_{2.5}$ during $2001-2010$ on all-cause mortality during $2011-2016$ in US zip codes. The estimated causal risk ratio $\E(Y(12))/\E(Y(9))$, comparing the mortality rate from $2011$ to $2016$ if all zip codes in the US had experienced an average PM$_{2.5}$ exposure of $12 \mu g/m^3$ (the primary annual National Ambient Air Quality Standard before 2024) versus $9 \mu g / m^3$ (the revised standard) during the period $2001-2010$, is approximately $1.034$. 

\begin{figure}
    \centering
    \includegraphics[width=\linewidth]{erfs_Apr2.png}
    \caption{Estimated exposure-response curves between long-term exposure to PM$_{2.5}$ during $2001-2010$ and all-cause mortality during $2011-2016$ among Medicare enrollees using seven different confounding adjustments.}
    \label{fig:ercs}
\end{figure}

\clearpage

\subsection{  Sensitivity analysis for the basis dimension}
Specifying the dimension of the spatial basis for confounding adjustment presents a bias-variance trade-off. A higher-dimensional basis can remove large-scale spatial variation in exposure, potentially isolating unconfounded variation, but at the cost of increased variance. \cite{dominici2004improved} established the theory behind this result in a temporal confounding setting and suggested a bootstrap-based approach for selecting the basis dimension. Although selection of the basis dimension used to construct $A_C$ is beyond the scope of the present work, we believe that the bootstrap procedure of \citet{dominici2004improved} may be readily adapted to our setting. More recently, \cite{keller2020a} proposed choosing the basis dimension for spatial confounding adjustment by fitting a model for the outcome as a function of covariates and the spatial basis, excluding exposure. 

We conducted a sensitivity analysis for the data application for each cutoff value $c = 6 \mu g/m^3, \ldots, 9 \mu g/m^3$. Specifically, we varied the degrees of freedom used in the thin plate spline and Graph Laplacian decompositions to assess how this choice influences the truncated exposure effect estimate. For the thin plate spline, we considered degrees of freedom ranging from 4 (the minimum allowable) to 10. For the Graph Laplacian, we varied the degrees of freedom from 1 to 10. In the main analysis, we selected basis dimensions of 4 and 3 for IV-TPS and IV-GL, respectively, which explained 22\% and 20\% of the variation in exposure.

\begin{figure}
    \centering
         \includegraphics[width = 0.47\linewidth]{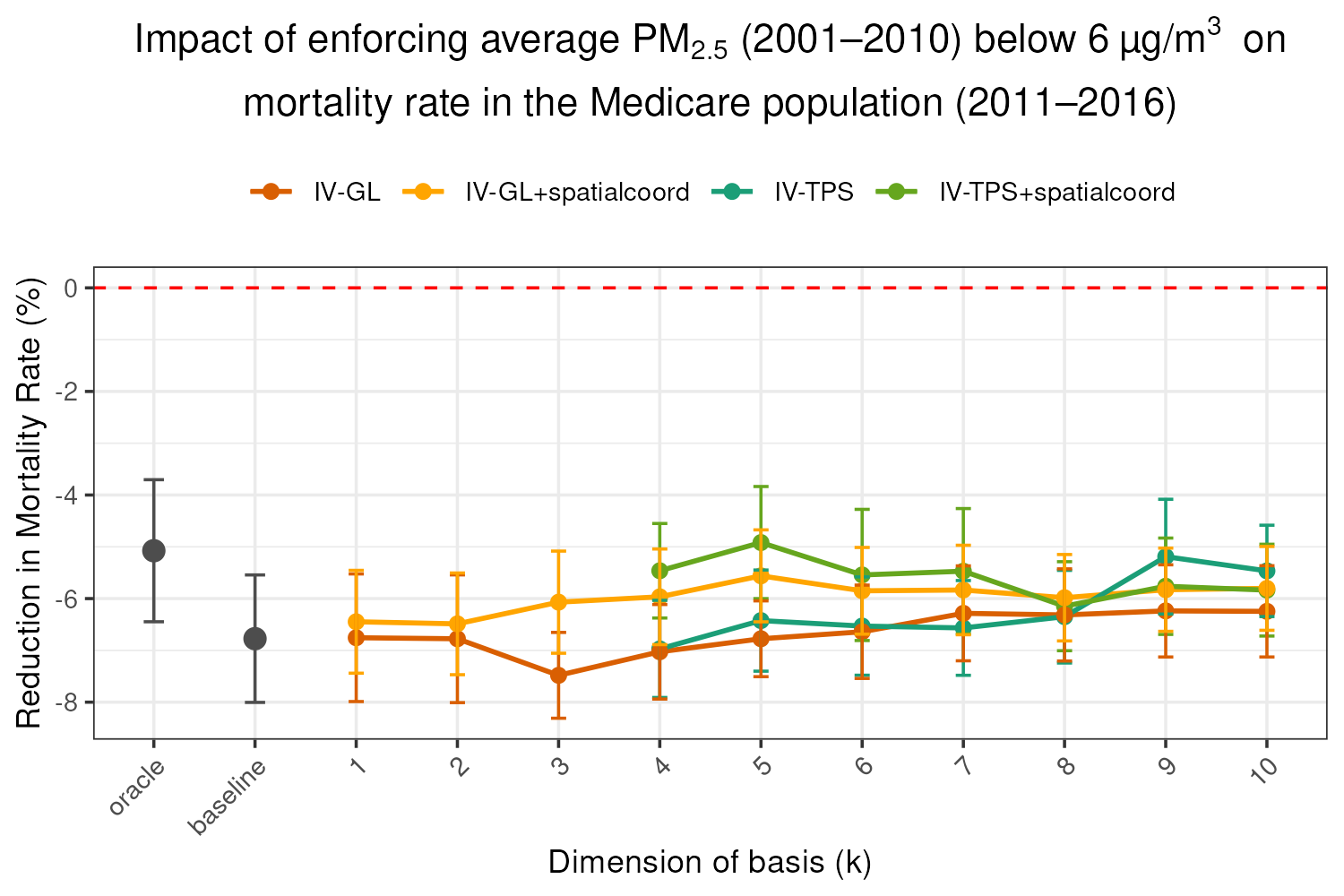}\includegraphics[width = 0.47\linewidth]{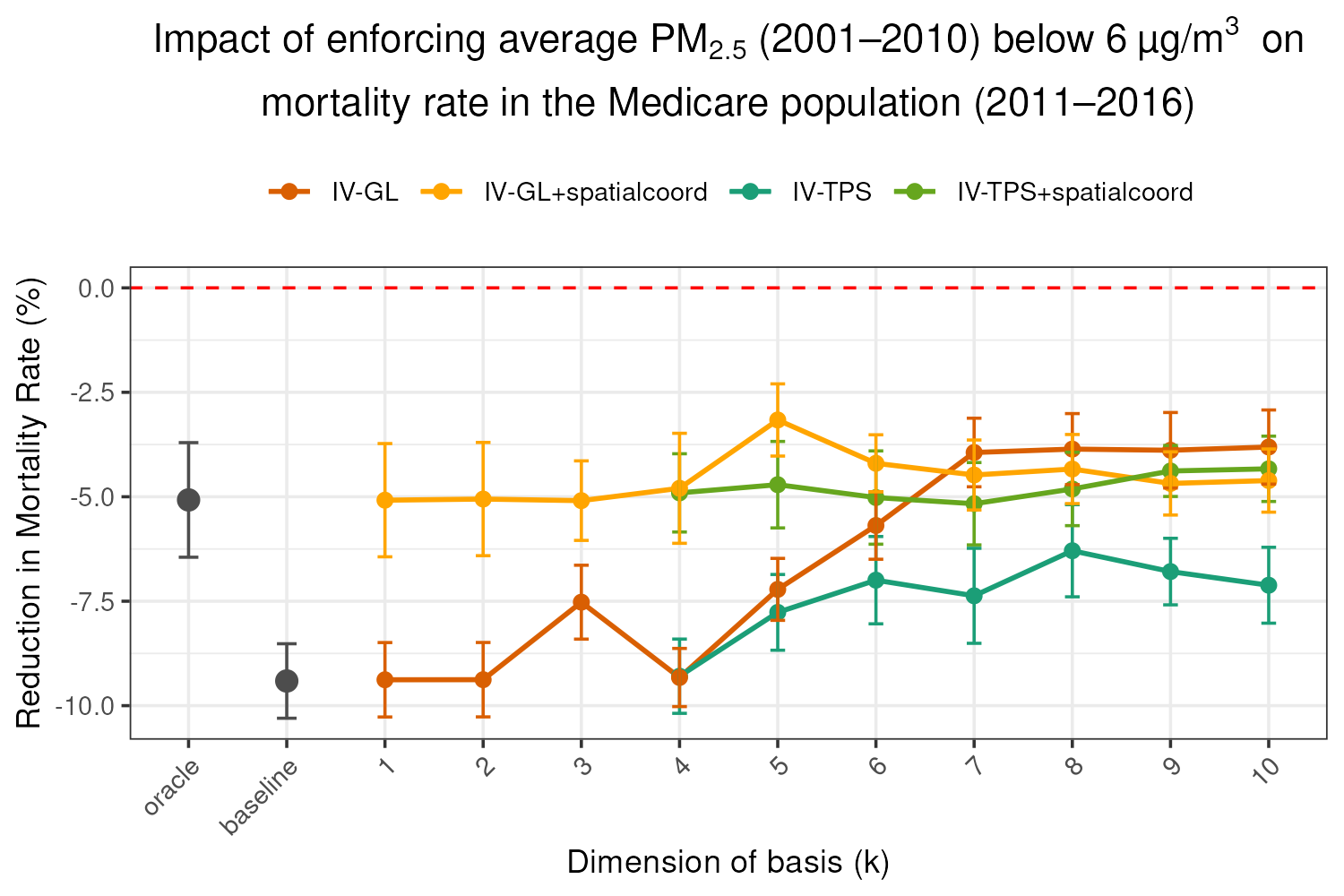}\\
         \includegraphics[width = 0.47\linewidth]{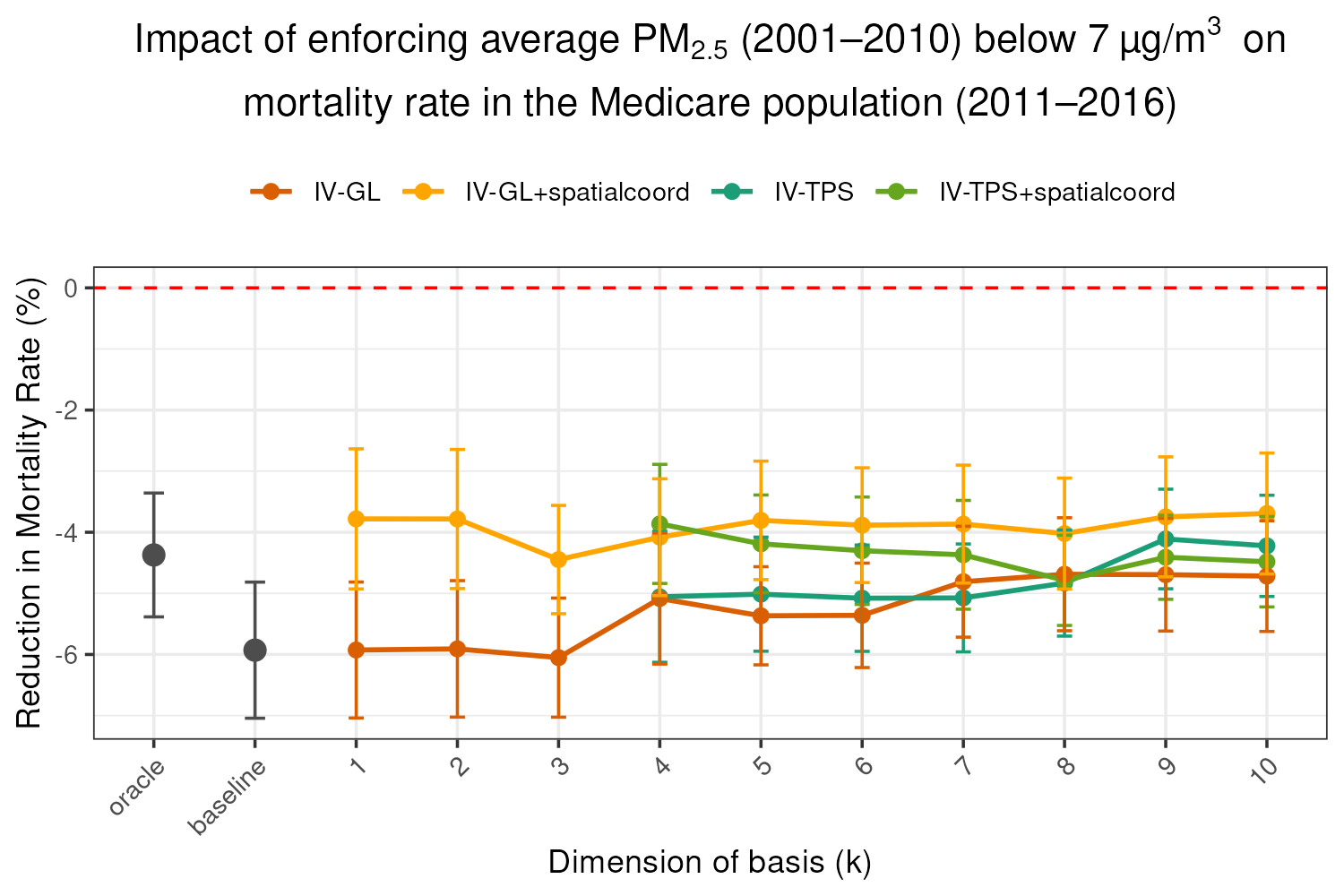}\includegraphics[width = 0.47\linewidth]{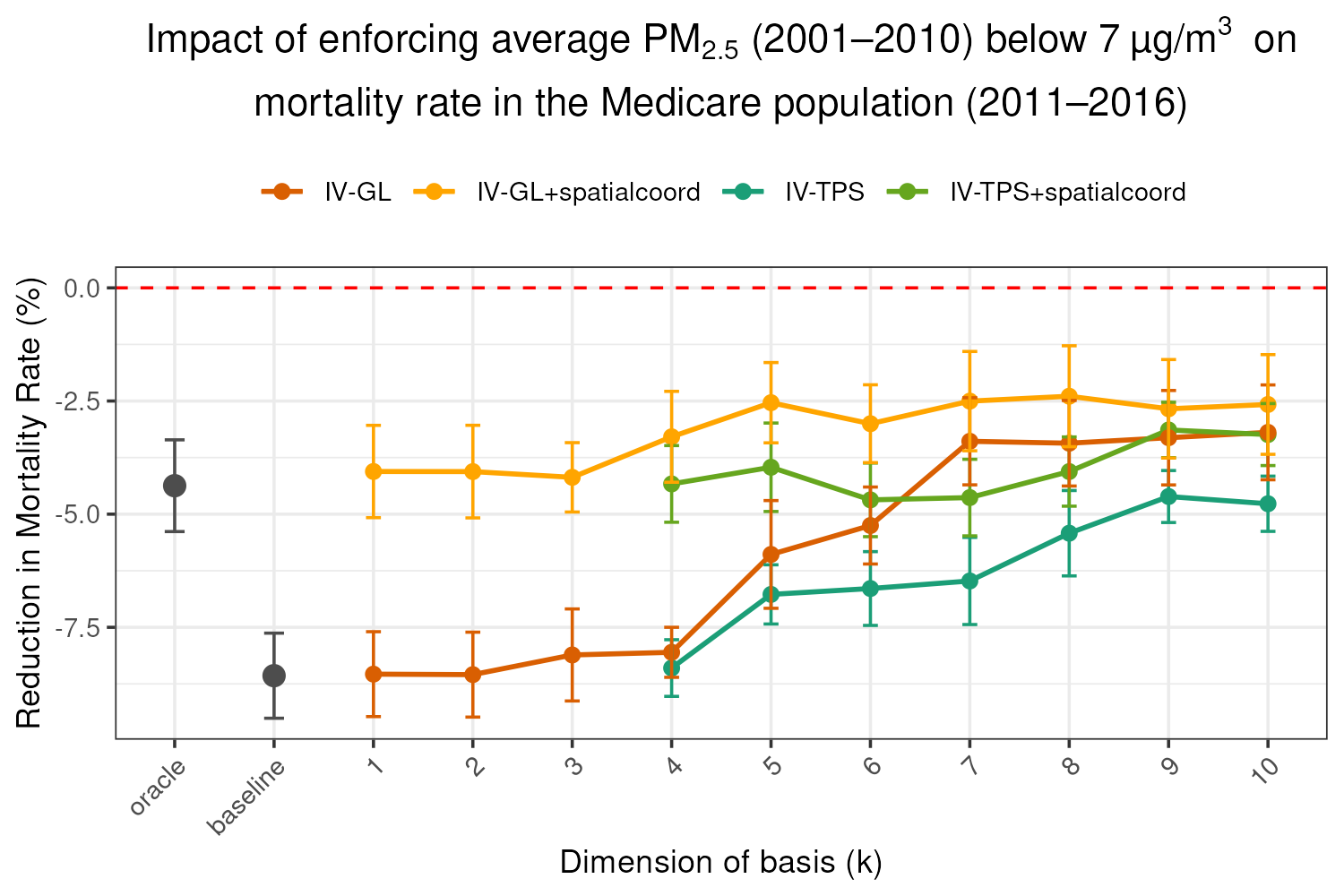}\\
         \includegraphics[width = 0.47\linewidth]{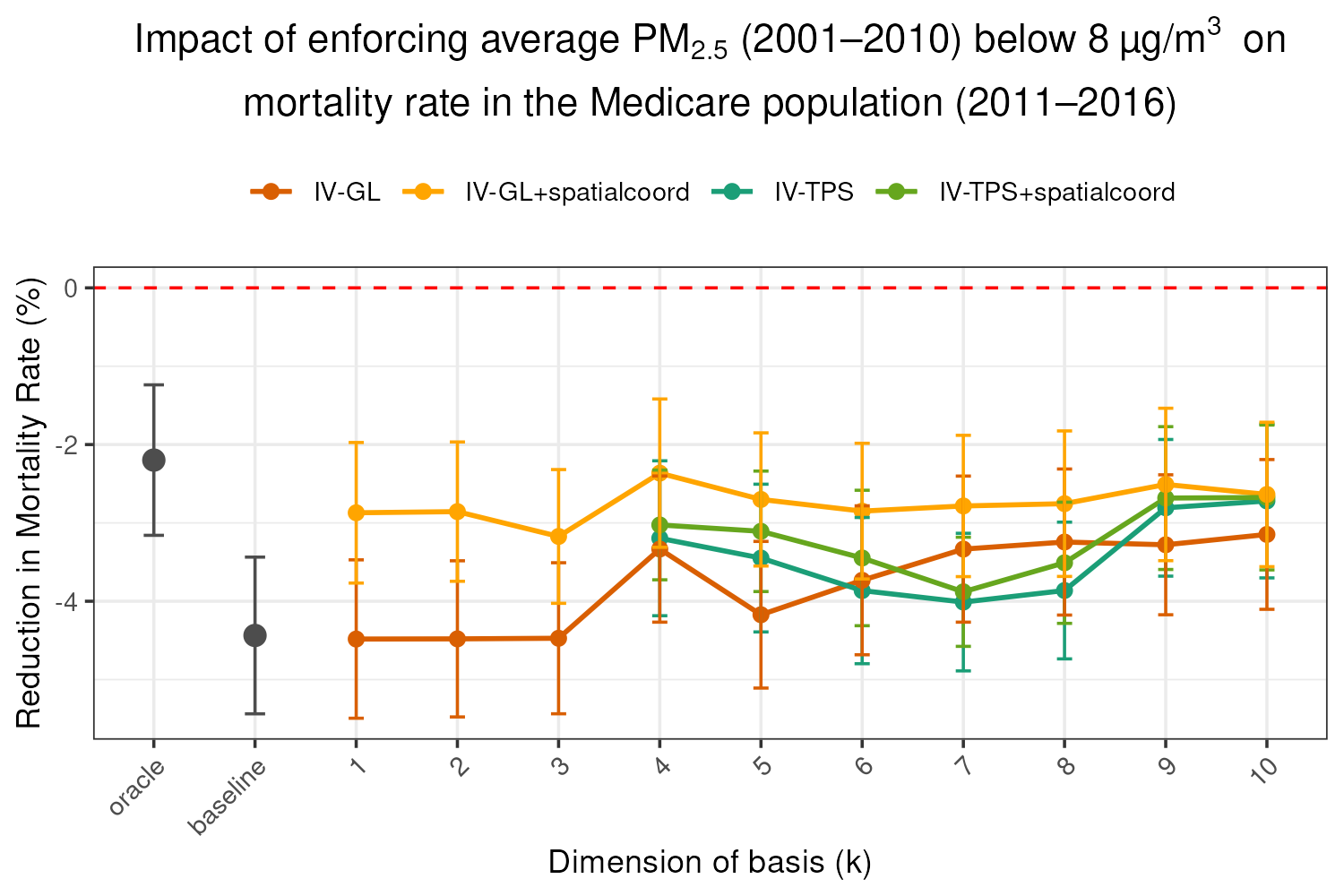}\includegraphics[width = 0.47\linewidth]{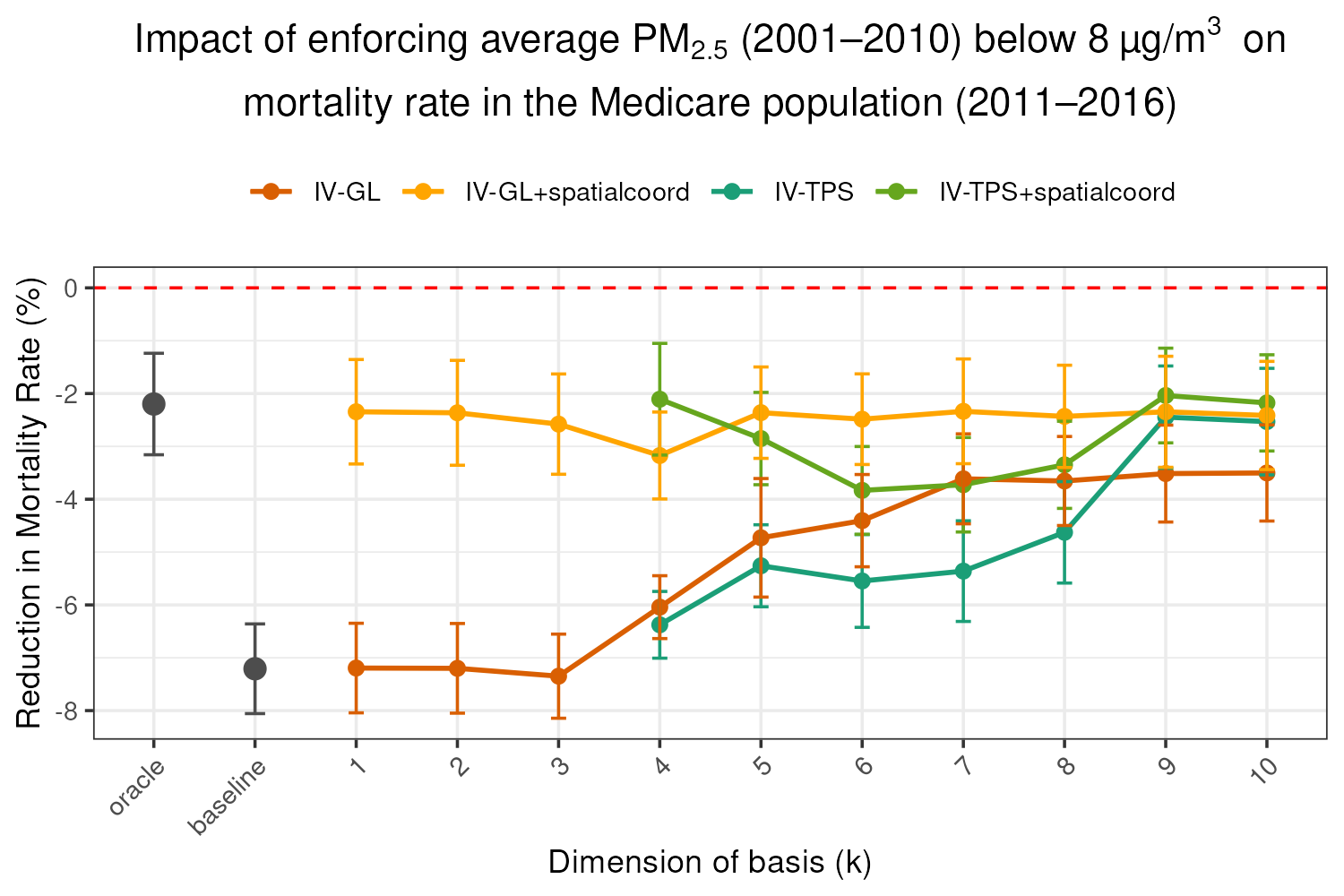}\\
         \includegraphics[width = 0.47\linewidth]{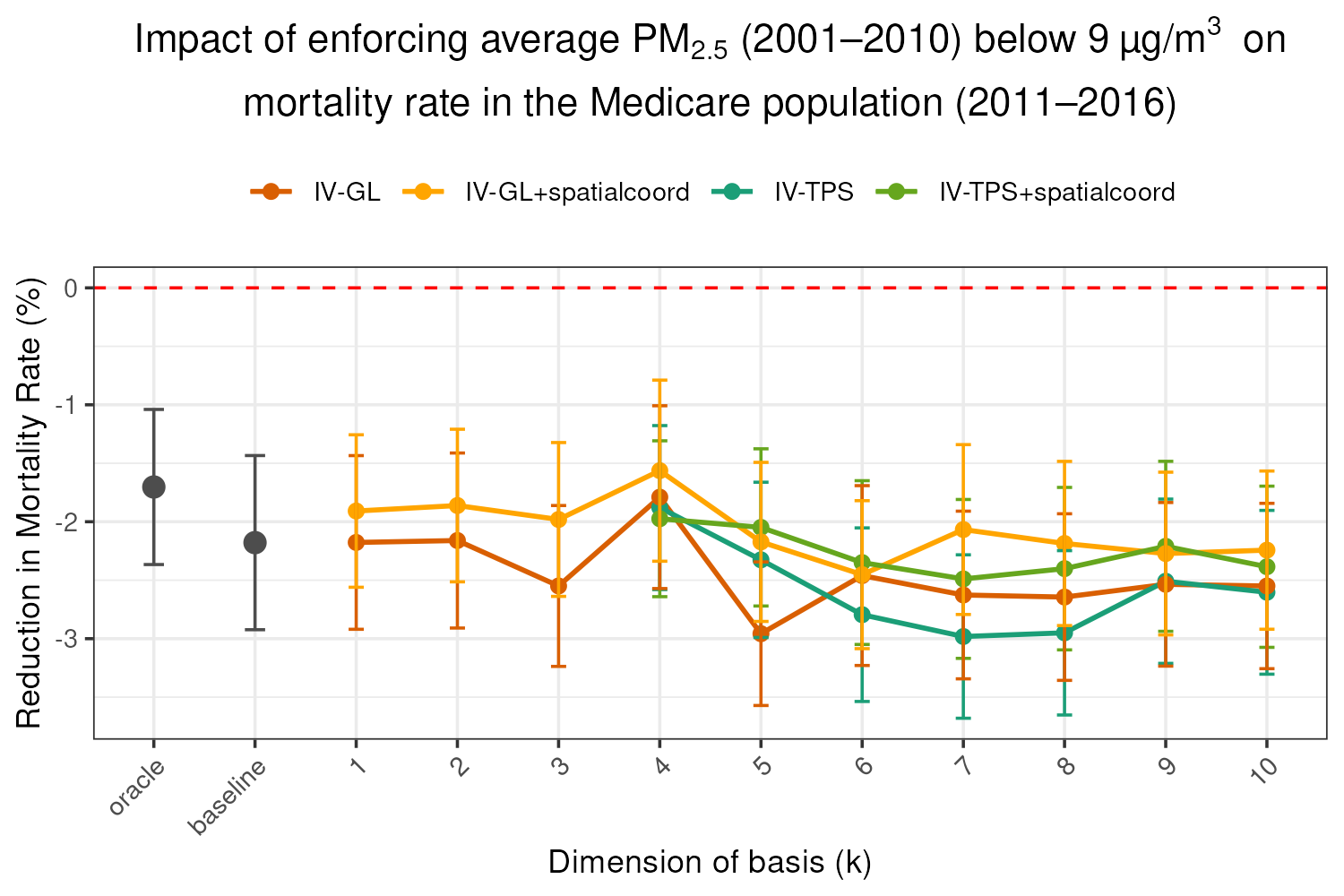}\includegraphics[width = 0.47\linewidth]{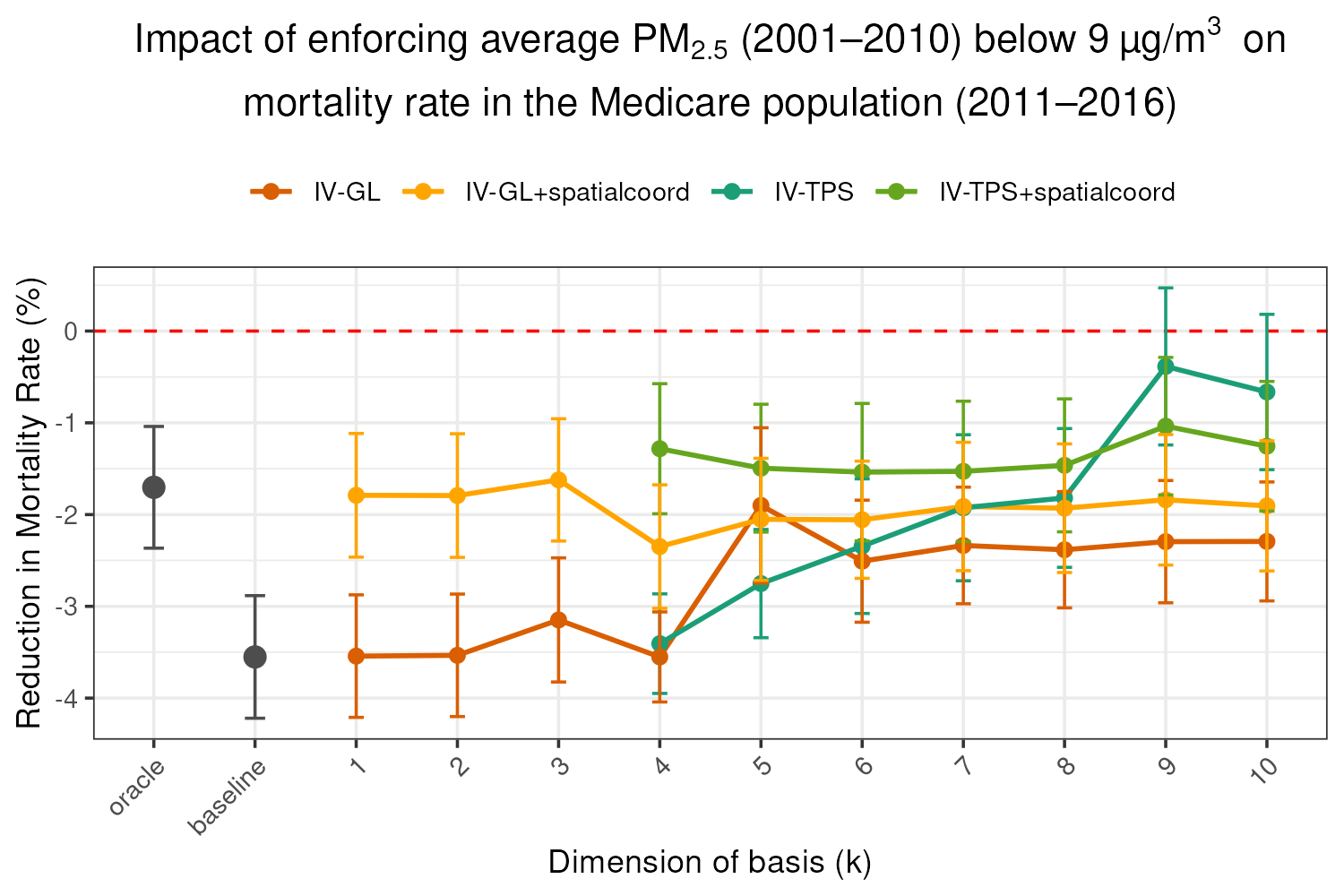}

\caption{\small{For cutoffs $6-9 \mu g/m^3$: a sensitivity analysis varying $k$, (i) the dimension of the thin plate spline used to create $A_C$ in the IV-TPS and IV-TPS+spatialcoord methods and (ii) the number of eigenvectors used to create $A_C$ in the IV-GL and IV-GL+spatialcoord methods. \textbf{Left panels}: the baseline and IV estimates adjust for $10$ measured covariates, as in the main analysis. \textbf{Right panels}: the baseline and IV estimates do not adjust for any measured covariates.}}
  \label{fig:sensitivity-analysis-1}
\end{figure}

Figure~\ref{fig:sensitivity-analysis-1} displays the truncated exposure effect estimates and their associated uncertainties as a function of the basis dimension. In the left columns, the baseline, oracle, and IV methods adjust for all $p = 10$ measured covariates as described in the main analysis. In the right columns, the baseline and IV methods do not adjust for any measured covariates. Specifically, the 
baseline approach does not adjust for any covariates beyond an intercept. IV-TPS only adjusts for the fitted values from an unpenalized thin plate spline regression of exposure on latitude and longitude, using $k = 4, \ldots, 10$ degrees of freedom. IV-GL only adjusts for the projection of exposure onto the smoothest $k$ eigenvectors of the Graph Laplacian, corresponding to the lowest nonzero eigenvalues, with $k = 1, \ldots, 10$.

Building on prior work documenting the bias–variance tradeoff \citep{dominici2004improved,  keller2020a}, we expected the IV methods to deviate from the baseline estimate and move toward the oracle estimate, with increasing variability, as the degrees of freedom $k$ used to construct $A_C$ increased. This pattern, however, does not consistently hold. We highlight three observations. 

\begin{enumerate}
    \item When the $p = 10$ additional covariates are included as measured confounders in both the outcome and conditional density models, effect estimates show only minor, nonmonotonic changes as $k$ varies (left columns of the figures). We attribute this to the fact that the measured covariates already capture substantial large-scale spatial information, so further adjustment through $A_C$ does not shift the estimates in a strictly monotonic fashion. 

    \item IV methods that also adjust for spatial coordinates (IV-TPS+spatialcoord and IV-GL+spatialcoord) are similarly stable across values of $k$, both with and without adjusting for the $p = 10$ measured covariates, likely for the same reason. 

    \item In contrast, in the non–covariate-adjusted scenarios (right columns), the IV-TPS and IV-GL estimates generally increase with $k$, moving from the baseline estimate (biased) to the oracle estimate (unbiased under strong assumptions) accompanied by increasing uncertainty. This pattern is consistent with the bias–variance tradeoff of \cite{dominici2004improved}. In fact, the least biased estimates arise when the degrees of freedom is closer to $k = 8,9,$ or $10$, suggesting that a larger $k$ may have been preferable in the main analysis. 
\end{enumerate}

\subsection{  An alternative strategy: spatial matching}
A reviewer noted that causal effects of interest could alternatively be estimated by matching on the spatial coordinates $S_i$; we agree that this would be an interesting direction for future work. Jointly matching on spatial coordinates and covariates is a complementary strategy aimed at adjusting for bias from unmeasured confounders that are continuous functions of spatial coordinates. Two existing works motivated by this idea are distance-adjusted propensity score matching \citep{papadogeorgou2019a}, designed for binary treatment, and conditional generalized propensity score-based spatial matching \citep{kim2022adjustment}, which extends this idea to accommodate continuous treatment settings.

Although neither work is directly applicable to estimating the truncated exposure effect $\E(Y(\min(A,c))$ we believe it would be possible to adapt these matching-based methods to our setting using the following pipeline:
\begin{enumerate}
    \item Estimate generalized propensity scores based on measured covariates, latitude, and longitude for the population with exposure values $> c$
    \item Estimate $\E(Y(c)|A > c)$ using existing methods that estimate the exposure-response curve with generalized propensity scores, such as \cite{wu2024matching}
    \item Estimate the truncated exposure effect $\E(Y(\min(A,c)) = \E(Y|A < c) \mathbb P(A < c) + \E(Y(c)|A > c)\mathbb P(A > c)$ using the estimate of $\E(Y(c)|A > c)$ from (2). 
\end{enumerate}
If Assumptions 6--9 hold, our IV framework suggests that one could instead match on values of $A_C$ and measured covariates instead of latitude, longitude, and measured covariates.
\end{document}